\documentclass[twocolumn, 10pt, pre, aps, showpacs, reprint, amsmath, amssymb, superscriptaddress, nofootinbib]{revtex4-2}

\usepackage[]{graphicx}

\begin{document}

\title{Complex nonlinear dynamics of polarization and transverse modes in a broad-area VCSEL}

\author{Stefan Bittner}
\email{stefan.bittner@centralesupelec.fr}
\author{Marc Sciamanna}
\affiliation{Chair in Photonics, CentraleSup\'elec, LMOPS, 2 rue Edouard Belin, Metz 57070, France}
\affiliation{Universit\'e de Lorraine, CentraleSup\'elec, LMOPS, 2 rue Edouard Belin, Metz 57070, France}

\begin{abstract}
Lasers can exhibit nonlinear and chaotic dynamics driven by the interaction of multiple lasing modes, and investigating the different scenarios of mode competition and bifurcations of the dynamics is of great interest on a fundamental level as well as in view of applications. We study the dynamics of a broad-area VCSEL in solitary continuous-wave operation with a comprehensive investigation of its polarization state, lasing spectra, near-field distributions and temporal dynamics. Fluctuations at the frequency of the birefringence splitting and other frequency components develop in a series of bifurcations. The bifurcations coincide with changes of the transverse lasing modes and/or the polarization state, demonstrating the importance of both the spatial and polarization degrees of freedom for the mode competition. In consequence, the inherent nonlinear dynamics of the broad-area VCSEL is significantly more complex than the dynamics of VCSELs with a single spatial mode.
\end{abstract}

\maketitle

\section{Introduction}
Semiconductor lasers (SCLs) can exhibit very rich and complex nonlinear dynamics due to the interaction of the optical field with the charge carriers \cite{OhtsuboBook2013}. While the resulting instabilities are detrimental to many applications, they are of considerable fundamental interest to study light-matter interaction and nonlinear dynamics \cite{Abraham1985, Weiss1988, Sciamanna2015}. Moreover, ultrafast complex and chaotic dynamics generated by SCLs have found applications in cryptography, microwave photonics, sensing and information processing \cite{Uchida2008, Qi2011, Sciamanna2015}. Chaotic dynamics is most often induced by delayed feedback \cite{Lenstra1985, Soriano2013}, however, the specific time scales introduced by the external cavity are strong limitations for security-sensitive applications \cite{Rontani2007}. Hence SCLs that are chaotic in solitary operation are of great interest and have the additional advantage of providing simpler and more compact systems. % paragraph

Instabilities and chaos in free-running SCLs can result from the interaction of multiple lasing modes via the active medium \cite{OhtsuboBook2013}. Examples are SCLs with several longitudinal modes \cite{Paoli1969, Furfaro2004}, vertical-cavity surface-emitting lasers (VCSELs) with two polarization modes \cite{Virte2013}, and SCLs with multiple transverse modes such as broad-area Fabry-Perot \cite{Fischer1996, Marciante1998, Scholz2008, Arahata2015} or polygonal lasers \cite{Ma2022a}. In particular highly-multimode edge-emitting SCLs have recently attracted attention since they can suppress filamentation \cite{Bittner2018a, Kim2022} and serve as low spatial-coherence light sources \cite{Kim2019, Cao2019} or ultrafast parallel random number generators \cite{Kim2021}. In contrast, the nonlinear dynamics of free-running broad-area VCSELs in continuous wave (cw) operation has not been investigated experimentally so far, which is the subject of this article. % paragraph

An important difference between edge-emitting SCLs and VCSELs is that the latter exhibit multiple polarization states. This additional degree of freedom can result in polarization switching, polarization-mode hopping \cite{Willemsen1999, Ackemann2001, Panajotov2008, Olejniczak2011, Virte2015}, nonlinear polarization dynamics \cite{MartinRegalado1996a, MartinRegalado1997b, Travagnin1996, vanExter1998, vanExter1998b, Ackemann2001, Sondermann2004, Olejniczak2011} and chaos \cite{Virte2013, Virte2013a}. The polarization and the dynamics of a VCSEL are intrinsically linked, and understanding them requires to account for the coupling of the optical polarization state to carrier reservoirs with different spins as well as their spin dynamics, for example in the framework of the spin-flip model \cite{SanMiguel1995}. Complex and chaotic polarization dynamics are reached via a sequence of polarization switching points and bifurcations, which has been discussed extensively for single transverse-mode VCSELs (called narrow VCSELs in the following) \cite{MartinRegalado1996a, MartinRegalado1997b, Virte2013, Virte2013a}, but not for broad-area VCSELs (BA-VCSELs). % paragraph

The presence of several transverse modes in BA-VCSELs  further increases the complexity. In this article we study the polarization state, dynamics and bifurcations of a BA-VCSEL as an example of a multimode laser exhibiting both spatial and polarization degrees of freedom. This study thus extends investigations into the dynamics of narrow VCSELs and broad-area edge-emitting SCLs alike. Previous experimental works on the dynamics of BA-VCSELs have studied the transient dynamics during short pulses \cite{Buccafusca1995, Buccafusca1996, Buccafusca1999, Giudici1998, Barchanski2003, Becker2004, Fuchs2007}, observing relaxation oscillations and the onset of mode competition and beating between transverse modes, whereas experiments in cw operation were restricted to the dynamics on long time scales \cite{Richie1994}. Numerical studies of BA-VCSELs \cite{Valle1995, MartinRegalado1997, Mulet2002, Mulet2002a, Hess1998, Hess2000, Babushkin2011} mostly concentrated on short pulses and transient dynamics as well, and a systematic investigation of the ultrafast dynamics and bifurcations of a BA-VCSEL under cw pump is still missing. Our main interests are understanding the typical frequency scales of the dynamics and the physical processes driving the bifurcations. % paragraph

We present a systematic study of the polarization state, lasing spectra, near-field profiles and temporal dynamics of a commercial BA-VCSEL in solitary cw operation. The laser exhibits a self-sustained complex nonlinear dynamics driven by a series of polarization-switching points and bifurcations unlike anything previously reported for narrow or broad-area VCSELs. The main frequency components are close to the birefringence and its (sub-) harmonics, but additional frequency components are generated at bifurcations. The bifurcations are related to changes of the transverse lasing modes, and it appears that the interplay of both spatial and polarization degrees of freedom via the charge carriers is driving the laser dynamics, resulting in a bifurcation sequence of unprecedented complexity. % paragraph

\section{Experimental setup} \label{sec:exp}
The laser under study is a BA-VCSEL with circular output aperture $15~\mu$m in diameter (Frankfurt Laser Company FL85-F1P1N-AC). The laser is pumped electrically in cw operation via a ring contact. All experiments are performed at room temperature, and the laser mount temperature is stabilized at $20^\circ$C with a thermo-electric cooler. % paragraph

\begin{figure}[tb]
\begin{center}
\includegraphics[width = 7 cm]{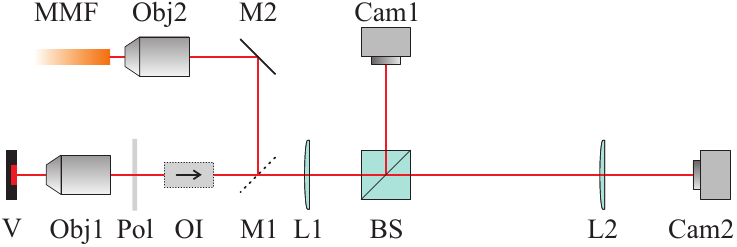}
\end{center}
\caption{Experimental setup. V: VCSEL; Obj1: $40\times$ objective (NA $= 0.6$); Pol: polarizer; OI: optical isolator; M1 and M2: mirrors; L1 (L2): lens with $200$~mm ($100$~mm) focal length; BS: beam splitter; Cam1 (Cam2): CCD camera in image plane (far-field) of VCSEL; Obj2: $20\times$ objective (NA $= 0.5$); MMF: multi-mode fiber}
\label{fig:setup}
\end{figure}

The experimental setup in Fig.~\ref{fig:setup} allows to measure near- and far-field images, optical spectra and temporal dynamics of the VCSEL. The laser emission is collimated with a $40\times$ microscope objective, and a lens with $200$~mm focal length is used to image the output facet on a CCD camera (Cam1). The pupil plane of the objective is imaged on a second CCD camera (Cam2) via an additional lens with $100$~mm focal length to measure the far-field images. A mirror on a flip mount (M1) behind the objective can divert the emission towards a fiber coupler with a graded-index multi-mode fiber (Thorlabs M116L02, NA $= 0.2$) which is connected to an optical spectrum analyzer (Anritsu MS9740A) or a fast photodetector (Newport 1544-B-50, DC-12 GHz). It was checked that bending or twisting the multimode fiber did not affect the measurement results in any way. An oscilloscope with $16$~GHz bandwidth (Tektronix DPO72340SX) is used to measure the time traces of the laser output power. For measurements of the laser dynamics, an optical isolator is inserted to avoid reflections from the fiber facet. Furthermore, the signal from the photodetector is amplified with a $29$~dB RF-amplifier (SHF S126A). Moreover a polarizer in a motorized rotation stage can be added in the beam path. % paragraph

\section{Polarization and optical spectrum} \label{sec:pol-spec}

\begin{figure}[tb]
\begin{center}
\includegraphics[width = 7 cm]{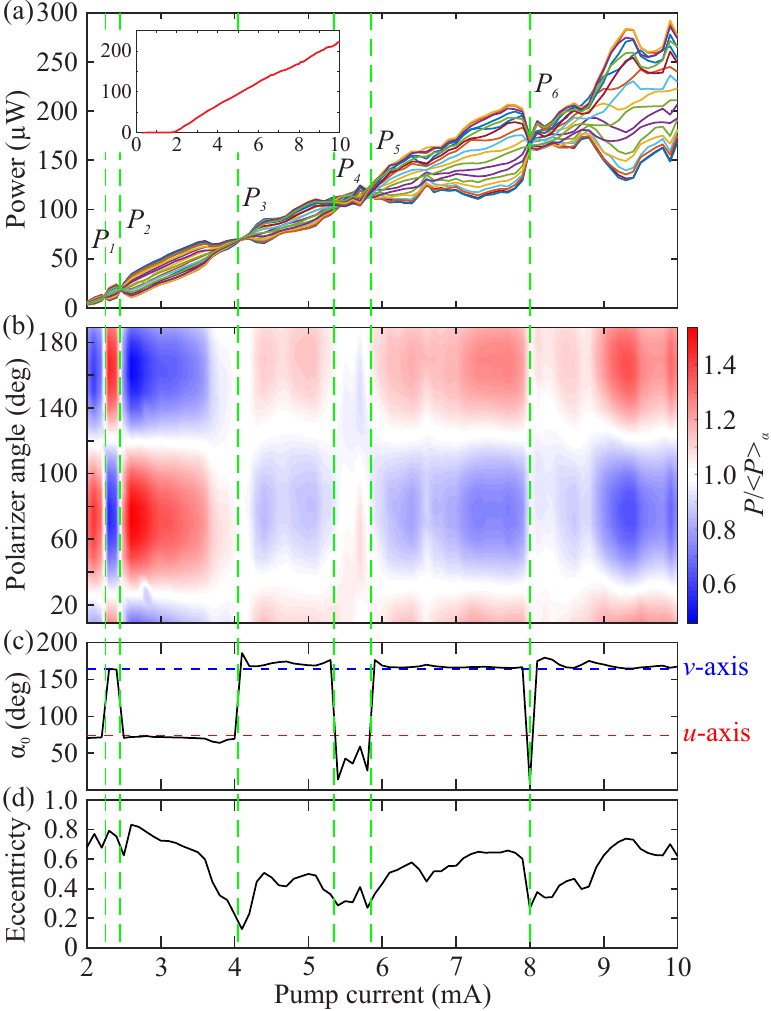}
\end{center}
\caption{LI-curve and polarization state. (a) LI-curves for $19$ different polarizer angles. Inset: LI-curve averaged over polarizer angles. (b) Power normalized by the average over all polarizer angles, $P / \langle P \rangle_\alpha$. (c) Angle $\alpha_0$ of major polarization axis. (d) Eccentricity $e$ of polarization state. The green dashed lines indicate the six polarization switching points, $P_1 = 2.25$~mA, $P_2 = 2.45$~mA, $P_3 = 4.05$~mA, $P_4 = 5.35$~mA, $P_5 = 5.85$~mA, and $P_6 = 8.0$~mA.}
\label{fig:LIcurve}
\end{figure}

First we investigate the polarization state of the VCSEL. The laser emission transmitted through a polarizer (see Fig.~\ref{fig:setup}) is measured with a power meter (Newport 818-IG) for different angles $\alpha$ of the polarizer axis with respect to the horizontal axis. The light-current (LI) curves for $19$ different polarizer angles $\alpha$ covering a range of $180^\circ$ are shown in Fig.~\ref{fig:LIcurve}(a). The inset shows the $\alpha$-averaged LI-curve with the threshold at $1.9$~mA. A total power slightly above $1$~mW is obtained for $10$~mA, though the measured power is four times lower due to the beam splitter and the polarizer. % paragraph

The level of power variations as function of the polarizer angle depends significantly on the pump current. The power normalized by its $\alpha$-average, $P_n(\alpha) = P(\alpha) / \langle P \rangle_\alpha$, is shown in false colors in Fig.~\ref{fig:LIcurve}(b). For a quantitative analysis, the normalized power is fitted by 
\begin{equation} \label{eq:polEll} P_n(\alpha) = 1 + c \cos[2 (\alpha - \alpha_0)] \end{equation}
for each pump current, where $\alpha_0$ is the major polarization axis and $c$ the power modulation amplitude. % paragraph

Figure~\ref{fig:LIcurve}(c) shows the major polarization axis $\alpha_0$ for different pump currents. We find the same two major orthogonal axes for practically all pump currents, called $u$ and $v$ axis in the following, with $\alpha_u \simeq 74^\circ$ and $\alpha_v \simeq 164^\circ$. The deviations from the $u$ axis found between $5.35$ and $5.85$~mA and at $8$~mA are attributed to problems with the fit due to noisy data in combination with a low modulation amplitude, whereas other measurements with the same device confirm that the polarization is indeed along the $u$ axis also in these current regimes. % paragraph

Figure~\ref{fig:LIcurve}(d) shows the eccentricity of the polarization given by $e = \sqrt{2 c / (1 + c)}$, where $e = 1$ indicates completely linear polarization along $\alpha_0$, whereas $e < 1$ indicates the emission is also partially polarized along the orthogonal axis (see Appendix~\ref{app:polStateAna} for details on the fit procedure and the meaning of the fit parameters). The eccentricity varies significantly with the pump current and is always below $1$ even just above threshold. In contrast, narrow VCSELs typically feature linear polarization at least in some current regimes and in particular close to threshold \cite{MartinRegalado1996a, MartinRegalado1997b, Travagnin1996, Virte2013, Virte2013a}. % paragraph

Figure~\ref{fig:LIcurve} exhibits six polarization switching points (PSPs) at which the dominant polarization axis switches from the $u$ to the $v$ axis or vice versa. The only exception is $P_6$ at which the polarization switches from the $v$ to the $u$ axis and directly back again, which may be caused by two very close PSPs that were not properly resolved. The PSPs are always accompanied by a dip in the eccentricity, though the dips are not necessarily very pronounced due to the finite current resolution of the measurement. This indicates that the lasing emission is shifted from one to the other polarization axis at the PSPs as shown in Fig.~\ref{fig:LIcurve}(a). In general, polarization switching can be caused either by a single lasing mode changing its polarization state or the (dis-) appearance of other lasing modes with different polarization states. Hence measuring the polarization state of individual lasing modes is needed to understand the cause of the PSPs. % paragraph

\begin{figure}[tb]
\begin{center}
\includegraphics[width = 7 cm]{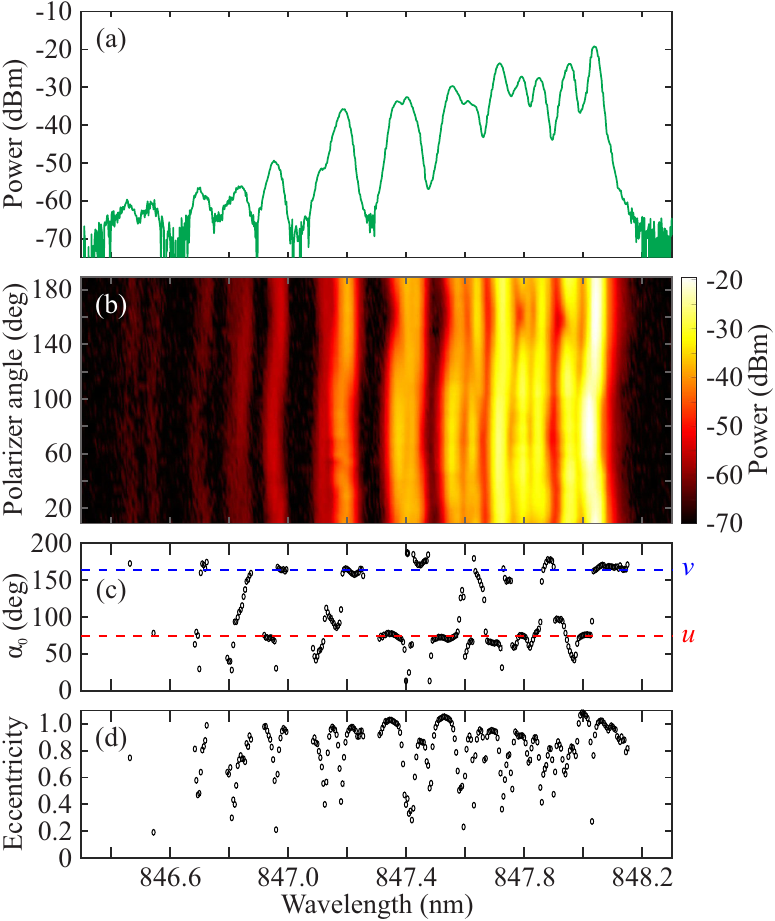}
\end{center}
\caption{Lasing spectrum at $6$~mA and polarization state. (a)~Spectrum measured without polarizer. (b)~Spectra as function of polarizer angle. (c) Angle $\alpha_0$ of major polarization axis and (d) eccentricity of polarization state as function of wavelength. The red (blue) dashed line in (c) indicates the $u$ ($v$) axis.}
\label{fig:spectraPol}
\end{figure}

Figure~\ref{fig:spectraPol}(a) shows the laser spectrum for $6$~mA measured without polarizer, which features a large number of partially overlapping lasing modes. As will be shown in the following, the spectrum features a number of different transverse modes, and each transverse mode additionally exhibits two different polarization states. The spectra as function of the polarizer angle $\alpha$ are shown in Fig.~\ref{fig:spectraPol}(b). The wavelength of individual peaks seems to vary with the polarizer angle, but in fact, these are pairs of near-degenerate modes with different polarization axes. Equation~(\ref{eq:polEll}) was fitted to all wavelength channels with sufficient signal, and Figs.~\ref{fig:spectraPol}(c) and (d) show the resulting major polarization axes and eccentricities as a function of wavelength. The values of $\alpha_0$ are concentrated along the $u$ and $v$ axes like in Fig.~\ref{fig:LIcurve}(c). The eccentricity depends significantly on the wavelength, and we observe peaks with $e \simeq 1$ at certain wavelengths surrounded by lower values of $e$. The reason for this behavior is the existence of near-degenerate modes with different polarization states: for each transverse mode, two modes with different polarization split by the birefringence $f_B \approx 9$~GHz are found. Since $f_B$ is of the same order as the spectral resolution of the spectrometer ($30$~pm), these modes overlap. The measured eccentricity of the polarization is hence low for wavelengths at which both modes contribute, and becomes maximal for wavelengths where only one mode contributes. Values of $e \approx 1$ demonstrate that the individual modes are highly linearly polarized\footnote{Values of $e > 1$ have no physical meaning but can appear due to the measurement noise.}. % paragraph

The same results are found for different pump currents as well. In conclusion, it appears from the data that all lasing modes are approximately linearly polarized along the two orthogonal axes $u$ and $v$ for all pump currents. While this is not necessarily unexpected for a VCSEL, it should be noted that other BA-VCSELs partially exhibit modes with elliptic and pump-dependent polarization state \cite{Molitor2015}. % paragraph

\begin{figure*}[tb]
\begin{center}
\includegraphics[width = 15 cm]{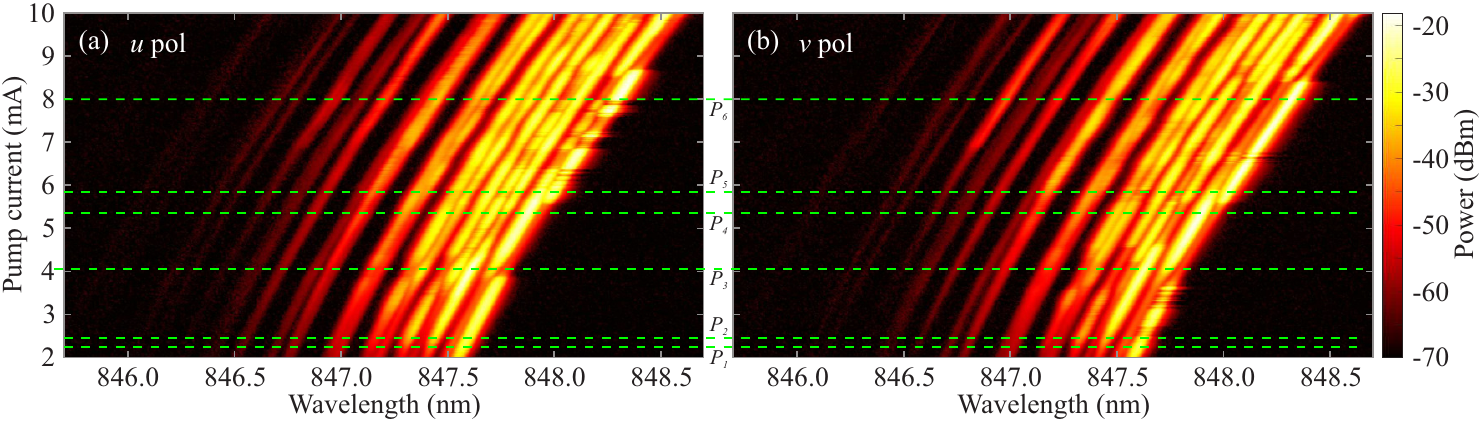}
\end{center}
\caption{Evolution of laser spectrum as function of pump current for (a) $u$ polarization and (b) $v$ polarization. The dashed green lines indicate the polarization switching points $P_{1-6}$ (cf.\ Fig.~\ref{fig:LIcurve}).}
\label{fig:spectraMainAxes}
\end{figure*}

The lasing spectra as function of the pump current measured with the polarizer along the $u$ and $v$ axes are shown in Fig.~\ref{fig:spectraMainAxes}. Even close to threshold, several transverse modes are observed in both polarizations. As the pump current is increased, more and more high-order modes (at short wavelengths) start lasing, but the low-order modes (at long wavelengths) continue to contribute significantly. Furthermore, a significant red-shift of the modes due to Joule heating is observed. % paragraph

Several factors contribute to the selection of transverse lasing modes. First, heating of the device changes the relative position of the gain maximum and the resonance wavelengths, and modifying the temperature of the laser mount results in a different evolution of the spectrum (not shown). Second, current crowding at the outer edge of the active area due to the ring contact favors high-order  modes of whispering-gallery type (see Section~\ref{sec:spatial} and Appendix \ref{app:NFevo}). However, these two effects cannot explain why for example the fundamental mode almost stops lasing around $3.9$~mA ($847.8$~nm), but comes back at $5.6$~mA ($848.0$~nm) only to stop lasing again around $8.7$~mA ($848.4$~nm), and other modes show similar behavior. This observation indicates that a complex mode competition via spatial hole burning takes place in addition to heating and current crowding effects. % paragraph

Apart from the gradual changes of the lasing spectrum with the pump current, we also observe erratic changes between different lasing mode configurations, for example in the range of $6$--$8$~mA in Fig.~\ref{fig:spectraMainAxes}(a). These fluctuations of the spectrum are shown in more detail in Fig.~\ref{fig:spectraFluc} in Appendix~\ref{app:specFluc}. Possible causes are an intrinsic multistability of the VCSEL or small thermal fluctuations of the environment. % paragraph

The presence of multiple lasing modes in both polarizations implies that the degree of polarization of the total lasing emission is low and explains the often low eccentricities in Fig.~\ref{fig:LIcurve}(d). The eccentricity of the total laser emission signifies the ratio of the power along the two polarization axes (see Appendix~\ref{app:polStateAna}), which gradually changes as the configuration of active lasing modes evolves. Hence, the PSPs indicated by the dashed green lines in Fig.~\ref{fig:spectraMainAxes} appear when the total power in the modes with $u$ polarization starts to exceed that of the modes with $v$ polarization (or vice versa), which results in dips in the eccentricity at the PSPs. While some of the PSPs such as $P_3$ at $4.05$~mA coincide with particular transverse modes turning off or on, these alone cannot explain the polarization switching since these mode changes usually happen symmetrically in both polarizations. This is an important difference to narrow VCSELs for which PSPs are caused either by individual lasing modes changing their polarization state (e.g., from linear to elliptic) or a second mode with different polarization turning on at the detriment of the first one \cite{Travagnin1996, MartinRegalado1996a, MartinRegalado1997a, MartinRegalado1997b, Balle1999, Ackemann2001, Sondermann2004, Olejniczak2009, Virte2013a}. For the BA-VCSEL considered here, in contrast, the individual lasing modes retain their linear polarization state, and PSPs result from gradual shifts of the power distribution among the modes in both polarization states. The relation of the PSPs to changes of the laser dynamics is discussed in Section~\ref{sec:discussion}. % paragraph

\section{Spatial structure and polarization} \label{sec:spatial}

\begin{figure}[tb]
\begin{center}
\includegraphics[width = 7 cm]{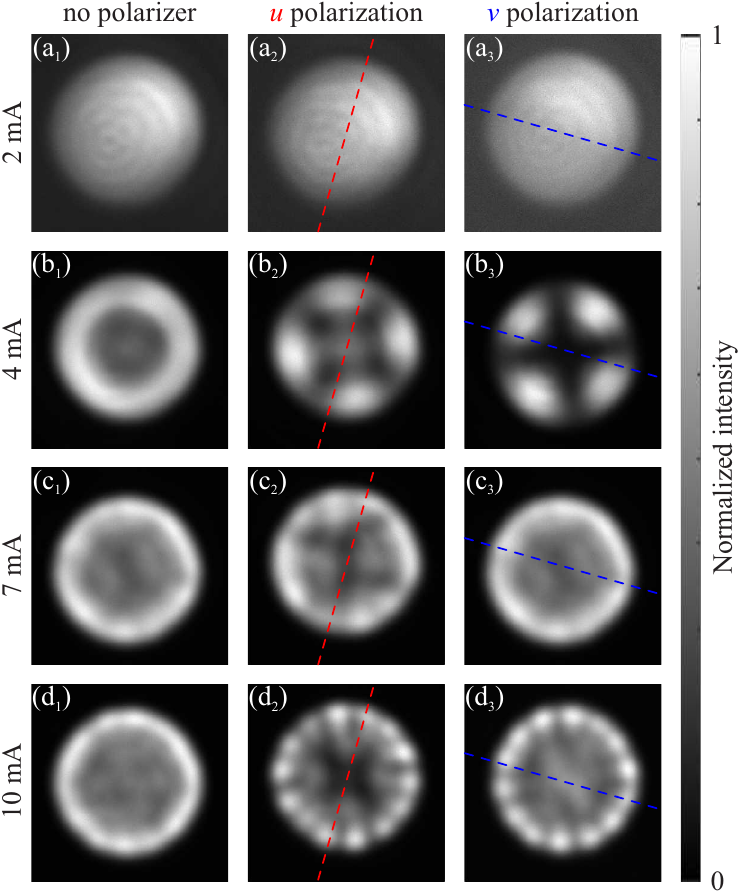}
\end{center}
\caption{Near-field images of the VCSEL with diameter $15~\mu$m for pump currents (a)~$2$~mA, (b)~$4$~mA, (c)~$7$~mA, and (d)~$10$~mA. The left column shows images without polarizer, the center column with polarizer along the $u$ axis (red dashed lines), and the right column with polarizer along the $v$ axis (blue dashed lines). The gray scale was normalized to the maximum of each image.}
\label{fig:NFimgs}
\end{figure}

Lasing in several transverse modes as observed for our BA-VCSEL implies that its near-field (NF) intensity distributions are formed by an incoherent superposition of modes. In the following we study the NF profiles to understand the different factors determining their shape, in particular current crowding and gain competition. % paragraph

Near-field images for different pump currents are shown in Fig.~\ref{fig:NFimgs}. Just above threshold [Fig.~\ref{fig:NFimgs}(a)], the NF profile is quite homogeneous, though slightly asymmetric with more intensity towards the upper right corner. When the pump current is increased, the images measured without polarizer (left column) show the concentration of the emission in an increasingly narrow ring at the boundary of the active region (see also Fig.~\ref{fig:NFsection} in Appendix~\ref{app:NFevo}). This is due to current crowding caused by the ring contact \cite{Degen1999} and partially explains the increasing strength of the high-order transverse modes observed in Fig.~\ref{fig:spectraMainAxes}. % paragraph

While the outer ring in the NF images looks rotationally symmetric, measurements with the polarizer (center and right columns in Fig.~\ref{fig:NFimgs}) reveal more details about the underlying transverse modes and their polarization. We label the different transverse modes with $(m, n)$, where $m$ ($n$) is the azimuthal (radial) quantum number and the intensity distribution exhibits $2m$ ($n$) maxima in azimuthal (radial) direction. It should be noted that the modes with $m > 0$ are two-fold degenerate (in addition to the two possible polarization states), where one mode has an intensity distribution $I \propto \cos^2(m \varphi - \varphi_0)$ (called cos-type mode) and the other one $I \propto \sin^2(m \varphi - \varphi_0)$ (called sin-type mode) with $\varphi$ the azimuthal angle and $\varphi_0$ an offset determining the orientation of the modes. For example at $4$~mA [Fig.~\ref{fig:NFimgs}(b)], the two NF images with polarizer both show an intensity distribution with four-fold symmetry, indicating lasing of the $(2, 1)$ mode. While the $(2, 1)$ mode is clearly dominant at $4$~mA, we observe in general a superposition of several different transverse modes. Hence, spatio-spectral measurements are needed to clearly identify the contributing modes (see Appendix~\ref{app:spatio-spec}). % paragraph

Figures~\ref{fig:NFimgs}(b)--(d) show two interesting features. First, the intensity distributions of the two polarization states are clearly anticorrelated, i.e., the intensity maxima for $u$ polarization are located at the minima of the $v$ polarization and vice versa. For example, Figs.~\ref{fig:NFimgs}(b$_2$) and (b$_3$) for $4$~mA both show a $(2, 1)$ mode, but one polarization exhibits the cos-type and the other one the sin-type mode. In addition, Fig.~\ref{fig:NFimgs}(b$_2$) shows a spot in the center stemming from a different transverse mode that is absent in Fig.~\ref{fig:NFimgs}(b$_3$). The situation is similar for $10$~mA [Figs.~\ref{fig:NFimgs}(d$_2$) and (d$_3$)], whereas the anticorrelation is less pronounced for $7$~mA [Figs.~\ref{fig:NFimgs}(c$_2$) and (c$_3$)]. Second, we find that the azimuthal positions of the maxima are unrelated to the polarization axes. % paragraph

\begin{figure}[tb]
\begin{center}
\includegraphics[width = 7 cm]{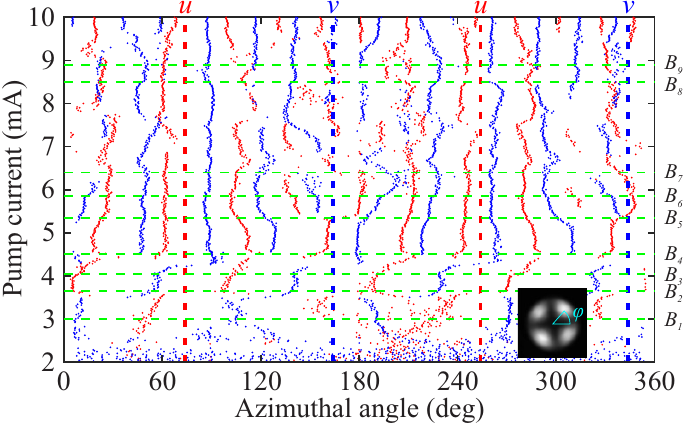}
\end{center}
\caption{Orientation of the near-field intensity distributions. The azimuthal positions of the maxima in the intensity distributions for $u$ ($v$) polarization are indicated as red (blue) dots. The vertical dashed red (blue) lines indicate the $u$ ($v$) polarization axis. The horizontal green dashed lines indicate the bifurcation points $B_{1-9}$ (see Fig.~\ref{fig:bifurcations}). Inset: near-field image for $4$~mA and $v$ polarization showing the definition of the azimuthal angle $\varphi$.}
\label{fig:imgOrient}
\end{figure}

A systematic evaluation of the azimuthal positions of the maxima in the NF images for $u$ and $v$ polarization is shown in Fig.~\ref{fig:imgOrient}. The number of maxima is often indicative of the $m$ of the dominant transverse modes. For example at $4.5$~mA, the number of maxima changes from four to eight because the $(2, 1)$ mode is superseded by the $(4, 1)$ mode. But while the azimuthal quantum number determines the angular spacing between maxima, $\pi / m$, the absolute position of the maxima given by $\varphi_0$ is \textit{a priori} undetermined for a VCSEL with perfect circular symmetry. % paragraph

In practice small perturbations or defects will pin the modes to a specific orientation. We note that for some currents, one maximum is aligned with a polarization axis, while for other currents this is not the case. Hence it appears that the birefringence axes do not determine the orientation of the modes. Small defects introduced during crystal growth or fabrication can potentially pin the modes \cite{Pereira1998}. Indeed, the positions of the maxima are mostly stable as long as the dominant mode does not change, but there are also examples of rapid changes of the orientation of a specific mode: at $3.8$~mA where the $(2, 1)$ mode is dominant, the maxima of the $u$ and $v$ polarizations suddenly interchange their positions, followed by a rapid rotation of the mode orientation till $4.5$~mA. A more detailed analysis of this rotation using spatio-spectral measurements is presented in Fig.~\ref{fig:modeRotation} in Appendix~\ref{app:spatio-spec}. So while structural defects most probably play a role in determining the orientation of the transverse modes, there seem to be also dynamical effects depending on the pump current since there is no azimuthal position that always displays a maximum and we cannot explain the rotation of the $(2, 1)$ mode around $4$~mA otherwise. % paragraph

Figure~\ref{fig:imgOrient} also confirms the anticorrelation of the spatial orientation of the two polarization states: the maxima for $u$ polarization are almost always found right between the maxima for $v$ polarization. The same effect has been previously observed in experiments with BA-VCSELs \cite{Pereira1998, Debernardi2002, Becker2004, Fuchs2007} and was attributed to spatial hole burning and a finite spin-flip relaxation rate in Refs.~\cite{MartinRegalado1997, MartinRegalado1997c}. It should be noted though that in spite of gain competition the fundamental mode $(0, 1)$ can lase in both polarizations at the same time (cf.\ Fig.~\ref{fig:spectraMainAxes}). In conclusion, our measurements strongly indicate that, in addition to spatial hole burning, the coupling of the optical polarization state with the dynamics of the carrier spins plays an important role in the selection of transverse modes and their polarization for the VCSEL considered here. % paragraph

\section{Dynamics and bifurcations} \label{sec:bifurcations}
To investigate the laser dynamics induced by the multimode competition, we measure the time traces of the laser output power using a fast photodetector and oscilloscope (see Section~\ref{sec:exp}). The total laser power is measured by aligning the input polarizer of the optical isolator between the $u$ and $v$ axes so both polarizations pass with equal amplitude. Since the power fluctuations are small (see Fig.~\ref{fig:TTstd} in Appendix~\ref{app:powFluc}), an RF-amplifier is used to increase the signal-to-noise ratio. Time traces of $20~\mu$s length are measured with a sampling rate of $50$~GSample/s for $2$--$10$~mA with a $0.01$~mA current step. The corresponding RF-spectra are obtained from the Fourier transform of the time traces, and are smoothed with a moving average over $5$~MHz. % pargraph

\begin{figure*}[tb]
\begin{center}
\includegraphics[width = 15 cm]{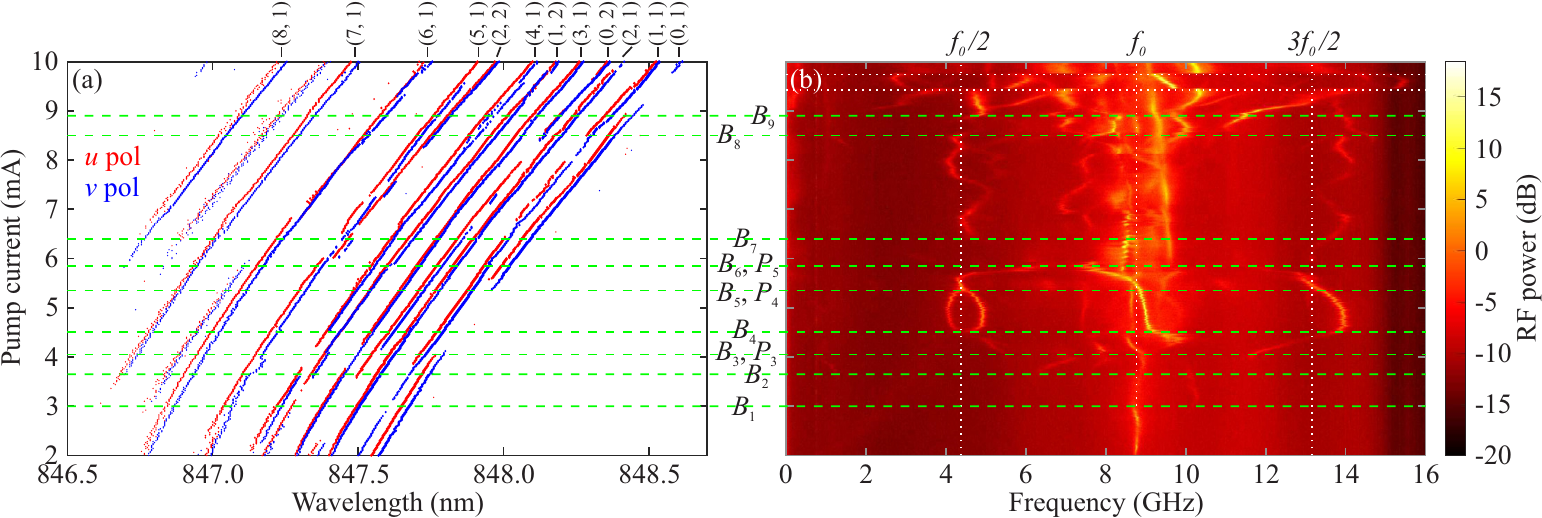}
\end{center}
\caption{Optical spectra, RF spectra and bifurcations. (a)~Optical spectra as function of current. The dots mark the peaks extracted from the spectra in Fig.~\ref{fig:spectraMainAxes}, where the dot size corresponds to the peak amplitude (in dBm), and modes with $u$ ($v$) polarization are shown in red (blue). The quantum numbers $(m, n)$ of the transverse modes are indicated above the figure. (b)~RF-spectra as function of current. The vertical white dotted lines indicate the basic fluctuation frequency $f_0$, its subharmonic $f_0/2$ and $3 f_0 /2$. The dashed green lines indicate the bifurcation points $B_{1-9}$, some of which coincide with the PSPs $P_{3-5}$. The horizontal dotted white lines indicate a sudden change of the dynamical state attributed to external perturbations.}
\label{fig:bifurcations}
\end{figure*}

The RF-spectra are shown together with the optical spectra for $u$ and $v$ polarization in Fig.~\ref{fig:bifurcations}. Each dot in Fig.~\ref{fig:bifurcations}(a) represents a peak above $-60$~dBm from Fig.~\ref{fig:spectraMainAxes}. The spectrum is organized in pairs of modes with the same transverse profile and orthogonal polarization. The quantum numbers $(m, n)$ obtained from the spatio-spectral measurements (see Appendix~\ref{app:spatio-spec}) are indicated above the spectrum. The frequency splitting between the different polarization states is $f_B \approx 9$~GHz and appears to be independent of the pump current (within the bounds of the measurement resolution). % paragraph

The RF-spectra in Fig.~\ref{fig:bifurcations}(b) show several bifurcations with qualitative changes of the observed frequency components (FCs). In the following we describe the nine major bifurcation points, $B_{1-9}$, which partially coincide with PSPs, and we detail the correlation of the bifurcations with changes in the optical spectrum and NF intensity distributions (see Fig.~\ref{fig:imgOrient}). Just above threshold a single FC at $f_0 \simeq 8.77$~GHz is found, which is in good agreement with the birefringence $f_B$ and presumably stems from beating of modes with different polarization states. At $B_1 = 3.0$~mA, the oscillation at $f_0$ becomes weaker, and another weak FC around $10.2$~GHz appears, but no significant changes in the optical spectrum are found. At $B_2 = 3.65$~mA, several FCs at $1$~GHz and between $6$ and $8$~GHz are created, whereas the FC at $10$~GHz disappears. Roughly at the same current the $(4, 1)$ modes vanish, the $(1, 2)$ modes appear in the optical spectrum, and the $(2, 1)$ modes with $u$ and $v$ polarization interchange their orientation. The bifurcation at $B_3 = 4.05$~mA sees the creation of two FCs at $5.5$ and $11.7$~GHz, while some of the other FCs disappear. $B_3$ coincides with the PSP $P_3$ and and the $(0, 1)$ modes turning off. All FCs including the one at $f_0$ undergo fast frequency changes as the current increases towards $B_4 = 4.5$~mA, and the $(2, 1)$ modes undergo a rapid rotation. At $B_4$ the FC at $4.5$~GHz splits into two components, and two FCs close to $f_0$ are found. In the optical spectrum, the $(4, 1)$ mode with $v$ polarization reappears at $B_4$ and starts to dominate the NF distribution (see Fig.~\ref{fig:imgOrient}). % paragraph

The next bifurcation at $B_5 = 5.35$~mA coincides with the PSP $P_4$. Furthermore, the $(0, 1)$ mode starts lasing again and the $(5, 1)$ becomes well visible in the NF images. At $B_5$ we see three FCs close to $f_0$, $f_0 / 2$ and $3 f_0 / 2$, which suggests period doubling of the birefringence beat note, but these frequencies quickly change as the current is increased further. Most of them disappear at $B_6 = 5.85$~mA, and only a few FCs in the range of $8$--$10$~GHz remain. $B_6$ correlates with the PSP $P_5$ and changes of the spectrum involving the $(1, 2)$ and the $(5, 1)$ modes. The FCs around $f_0$ continue to dominate the RF-spectrum. In addition, two FCs near $f_0/2$ and $3 f_0/2$ reappear at $B_7 = 6.4$~mA, which roughly coincides with the disappearance of the $(1, 1)$ mode with $u$ polarization and shifts of the $(2, 2)$ and $(5, 1)$ modes. At $B_8 = 8.5$~mA, a FC exactly at $f_0$ as well as further FCs in the vicinity are created, whereas the FC around $f_0 / 2$ vanishes. At the same current, the $(0, 2)$ modes and the $(1, 1)$ mode with $u$ polarization start lasing again. The FC at $f_0$ vanishes again at $B_9 = 8.9$~mA, whereas other FCs including one at $4.9$~GHz appear. Furthermore the $(0, 1)$ mode with $u$ polarization disappears once more at $B_9$. In addition we find sudden changes of the RF-spectra at $9.43$ and $9.74$~mA, which we attribute however to instabilities of the lasing spectrum (cf.\ Fig.~\ref{fig:spectraMainAxes} and Appendix~\ref{app:specFluc}). % paragraph

In summary, we find a complex cascade of bifurcations, where most of the observed FCs are around $f_0 \simeq 8.77$~GHz, which seems to be the birefringence, as well as around $f_0/2$ and $3 f_0 / 2$. This indicates period doubling of a self-pulsing dynamics driven by the birefringence frequency. However, we also often find several FCs around $f_0$ which are incommensurable, so other effects besides period doubling must be involved. Both the number of FCs and their frequencies increase on average with the pump current. Measurements with decreasing pump current showed no sign of hysteresis. While we cannot give a comprehensive explanation of all bifurcations, we can gain some physical insights by comparing our findings with previous studies of VCSEL dynamics in the following section. % paragraph

\section{Discussion} \label{sec:discussion}
The dynamics of narrow VCSELs has been intensively investigated with particular attention to the role of polarization-switching points \cite{MartinRegalado1996a, MartinRegalado1997b, vanExter1998, vanExter1998b, Ackemann2001, Sondermann2004, Olejniczak2011, Virte2013, Virte2013a}. While polarization switching can be induced by changes of the relative gain of two polarization modes, for example due to thermal effects, it can also be caused by bifurcations in which one polarization state loses its stability in favor of a different state \cite{MartinRegalado1996a, MartinRegalado1997b}. A typical scenario of bifurcations found for narrow VCSELs \cite{MartinRegalado1996a, MartinRegalado1997b, Travagnin1996, Virte2013, Virte2013a} is that the initial linear polarization state loses stability and is replaced by an elliptical polarization state with stationary dynamics as the current is increased. Further bifurcations lead to self-pulsations of the elliptical state and even chaotic dynamics before the laser stabilizes again for even higher currents. A spin-flip type model \cite{SanMiguel1995} for the VCSEL dynamics is required to explain this scenario \cite{MartinRegalado1996a, MartinRegalado1997b, Virte2013a}. The BA-VCSEL studied here, however, exhibits a more complicated series of bifurcations with a larger number of PSPs and without restabilization of the dynamics. An important difference is that the PSPs leading to unstable dynamics in narrow VCSELs are attributed to individual modes changing their polarization state, whereas for our BA-VCSEL each mode retains its polarization state, and the PSPs are instead caused by a gradual redistribution of the power among different transverse	 modes. So lasing in several transverse modes results in a significantly more complex evolution of polarization and dynamics compared to narrow VCSELs with different mechanisms causing the bifurcations and PSPs. % paragraph

The frequencies found in the dynamics of a system are a key property that allows insight into the responsible physical processes. While the typical frequency scales found here are given by the birefringence and its (sub-) harmonics, most FCs and their evolution with the pump current cannot be explained in a simple way. Besides the birefringence, the relaxation oscillation frequency often plays an important role \cite{Olejniczak2011, Virte2013, Buccafusca1995}, but a FC with the square-root dependence on the pump current characteristic for relaxation oscillations is not discernible in Fig.~\ref{fig:bifurcations}(b). Beating between transverse modes was predicted \cite{MartinRegalado1997} and demonstrated experimentally \cite{Buccafusca1995, Buccafusca1996} for BA-VCSELs, however, these beat frequencies are very high for our VCSEL and beyond the bandwidth of the photodetector. In addition, the frequencies of beat notes should exhibit little dependence on the pump current, whereas many of the FCs in Fig.~\ref{fig:bifurcations}(b) change rapidly with the current. Furthermore, fluctuations on the scale of $10$--$30$~GHz caused by mode competition were observed in BA-VCSELs \cite{Buccafusca1995, Buccafusca1996}, however, these do not match the FCs observed here either. Moreover, almost all investigations of the dynamics of BA-VCSELs only considered short pulses, observing relaxations oscillations and the onset of mode competition \cite{Buccafusca1995, Buccafusca1996, Buccafusca1999, Giudici1998, Barchanski2003, Becker2004, Fuchs2007}, however, it is not clear how and with which frequency scales mode competition persists in cw operation. We thus cannot clearly identify the origin of all the FCs observed here, but it appears very likely that they are due to processes different from those previously observed for BA-VCSELs. % paragraph

Another type of dynamics found for narrow VCSELs is polarization-mode hopping in regimes of bistability, typically found near PSPs \cite{Willemsen1999, Ackemann2001, Panajotov2008, Olejniczak2011, Virte2015}, where the VCSEL fluctuates erratically between two different polarization states. These fluctuations have time scales ranging from the nanosecond to the second regime and normally become slower with increasing pump \cite{Willemsen1999, Panajotov2008}. However, measurements of the time traces with the polarizer do not show any polarization-mode hopping (see Appendix~\ref{app:powFluc}). Moreover, the frequency scales that we observe and their evolution with the current are very different from what is found for polarization-mode hopping; in particular no slowing down of the dynamics is observed. A BA-VCSEL might also feature hopping between different transverse modes \cite{Richie1994}, however, the time scales of this dynamics are again much slower than what we measure. % paragraph

In summary, the frequencies of the nonlinear dynamics observed in our laser are unlike those previously found for both narrow and broad-area VCSELs and defy any simple explanation. We believe they result from a cascade of Hopf bifurcations, which can create frequencies similar to but different from the birefringence \cite{Virte2013a}. Detailed theoretical studies are needed to understand the bifurcations of BA-VCSELs, but some qualitative insights can be gained from the experimental data (see Section~\ref{sec:bifurcations}). It appears that changes in the transverse lasing modes are of great importance since they are found near most of the bifurcation points. Some bifurcations also coincide with polarization switching, but PSPs clearly do not play the same role as in the dynamics of narrow VCSELs. Nonetheless the predominance of the birefringence frequency in the dynamics demonstrates the importance of polarization effects. While the bifurcations are related to the transverse mode changes driven by spatial hole burning, the coupling of the optical polarization state with the spin dynamics of the carriers also plays an important role as discussed in Section~\ref{sec:spatial}. Thus considering the spatial and polarization degree of freedom of the optical field in conjunction with the spin dynamics of the carriers \cite{MartinRegalado1997, Hofmann1997a, Mulet2002a} is required to explain the complex mode competition and nonlinear dynamics we observe, though thermal effects must not be neglected either in cw operation \cite{MartinRegalado1997a}. % paragraph

\section{Summary and outlook} \label{sec:conclusion}
In our systematic spatial, spectral, dynamical and polarization-resolved study of a commercial BA-VCSEL with cw pumping we find a complicated series of bifurcations and polarization switching points. The bifurcations are closely related to the evolution of the lasing spectrum which is governed by transverse mode competition. Such complex nonlinear dynamics is not often found for free-running semiconductor lasers, which makes it interesting for chaos-based applications \cite{Sciamanna2015}. Measurements with a second device of the same type show the same qualitative behavior regarding mode competition, polarization state and polarization switching, spatial structure, frequency components in the RF-spectra as well as bifurcation sequence. Although a detailed comparison of the dynamics of different devices will be published in a future article, this already indicates that the complex dynamics reported here is a generic behavior for this kind of BA-VCSEL. % paragraph

Our results agree qualitatively with predictions by the spin-flip model \cite{SanMiguel1995}: first, we observe oscillation frequencies around the birefringence frequency and the creation of new frequency components in a sequence of bifurcations \cite{Virte2013a}. Second, we observe that the intensity distributions of the two polarization states are spatially anticorrelated, which clearly shows the importance of the carrier-spin dynamics in the mode competition dynamics \cite{MartinRegalado1997, MartinRegalado1997c}. However, the bifurcation scenario of the BA-VCSEL is different from and more complex than those found for narrow VCSELs \cite{MartinRegalado1996a, MartinRegalado1997b, Virte2013, Virte2013a} because of the presence of several transverse modes. The dynamics in cw operation is also different from the dynamics previously observed for BA-VCSELs in pulsed operation \cite{Buccafusca1995, Buccafusca1996, Buccafusca1999, Giudici1998, Barchanski2003, Becker2004, Fuchs2007}. In short, the nonlinear dynamics of our BA-VCSEL seems to be unlike any discussed so far in the literature. Even though models of BA-VCSELs including polarization effects and carrier-spin dynamics exist \cite{MartinRegalado1997, Hofmann1997a, Mulet2002a}, the resulting bifurcation scenarios have not been investigated so far, and it remains to be seen if existing models are capable of explaining our experimental findings. % paragraph

Further experiments are needed to better characterize the temporal and polarization dynamics of the BA-VCSEL \cite{Molitor2016, Ploeschner2022} and determine its possibly chaotic properties. Our results in Fig.~\ref{fig:bifurcations} demonstrate that the nonlinear dynamics including self-pulsation and higher-order bifurcations like period doubling originates from the interplay of many modes with different spatial and polarization properties. We observe a number of frequency components up to the bandwidth of our detection system, and even higher frequency components due to transverse mode beating are expected \cite{Buccafusca1995, Buccafusca1996, MartinRegalado1997} to further enhance the bandwidth of the nonlinear dynamics. It is therefore conceivable that chaotic dynamics with higher bandwidth and greater complexity than for SCLs with optical feedback \cite{Bouchez2019} can be attained with BA-VCSELs. In addition, more comprehensive studies of how the nonlinear dynamics and the underlying bifurcations depend on the temperature, optical injection or feedback would be important. Finally, the complex dynamics of the BA-VCSEL could be harnessed for applications like reservoir computing \cite{Porte2021} since it has been found that reservoir computing performance is improved by polarization-mode competition in narrow VCSELs \cite{Vatin2019}. % paragraph

Moreover it would be interesting to study asymmetric VCSELs with chaotic ray dynamics. A non-circular cavity cross section will break the spatial degeneracy of the transverse modes and modify the polarization states as well \cite{Chen2007b, Babushkin2008}, which should impact the mode competition and thus the lasing dynamics. While the spatial structure formation in asymmetric BA-VCSELs has been studied on the static level \cite{Huang2002, Chen2003a, Brejnak2021}, there are no investigations of their dynamical properties so far. An additional interest is the development of highly-multimode VCSELs with low spatial coherence. Experiments with edge-emitting asymmetric microcavity lasers have already shown the great potential of asymmetric cavities to manipulate the dynamics and spatial coherence of SCLs \cite{Bittner2018a, Kim2019, Cao2019, Kim2022}. % paragraph

\begin{acknowledgments}
S.~B.\ thanks Hui Cao, Kyungduk Kim, Ortwin Hess and Stefano Guazzotti for fruitful discussions. S.B.\ and M.S.\ thank Mario Fernandes for technical support.
S.~B.\ and M.~S.\ acknowledge support for the Chair in Photonics from Minist\`ere de l'Enseignement Sup\'erieur, de la Recherche et de l'Innovation; R\'egion Grand-Est; D\'epartement Moselle; European Regional Development Fund (ERDF); Metz M\'etropole; GDI Simulation; CentraleSup\'elec; Fondation CentraleSup\'elec. 
\end{acknowledgments}

\appendix

\section{Polarization state analysis} \label{app:polStateAna}

\begin{figure}[tb]
\begin{center}
\includegraphics[width = 7 cm]{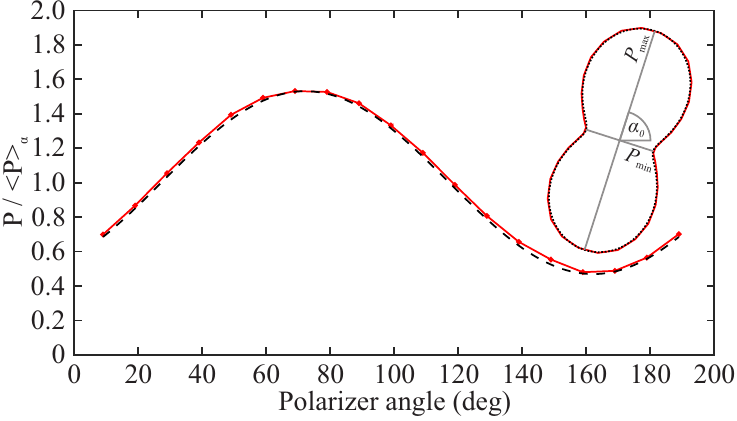}
\end{center}
\caption{Normalized power of the VCSEL transmitted through the polarizer as function of the polarizer angle $\alpha$ for $2.6$~mA. The measured power is shown in red, the fit of Eq.~(\ref{eq:polEllFit}) as black dashed line. Inset: the same data as polar plot, where $\alpha_0$ is the angle of the major axis.}
\label{fig:polEllFit}
\end{figure}

The setup in Fig.~\ref{fig:setup} allows to measure the laser transmission $P(\alpha)$ through a polarizer as function of the angle $\alpha$ of the polarizer axis with respect to the horizontal axis. The measured transmission as function of $\alpha$ and normalized to the $\alpha$-averaged power is well described by the formula
\begin{equation} \label{eq:polEllFit} P(\alpha) / \langle P \rangle_\alpha = 1 + c \cos[2 (\alpha - \alpha_0)] \end{equation}
where $c \in [0, 1]$ is the power modulation amplitude and $\alpha_0$ the angle of the major axis. The example in Fig.~\ref{fig:polEllFit} shows the measured power as function of the polarizer angle for $2.6$~mA (red) and the fit of Eq.~(\ref{eq:polEllFit}) (black dashed line), which shows very good agreement. The fit parameters are $\alpha_0 = 72.2^\circ$ and $c = 0.531$. % paragraph

The interpretation of the fit parameters depends on whether the laser light is fully or partially polarized, which is not evident since the measurement does not yield the complete set of Stokes parameters. For fully polarized light, Eq.~(\ref{eq:polEllFit}) describes an elliptical polarization state with major axis $\alpha_0$ and eccentricity
\begin{equation} e = \sqrt{1 - P_\mathrm{min} / P_\mathrm{max}} = \sqrt{2 c / (1 + c)} \, , \end{equation}
where $P_\mathrm{max}$ ($P_\mathrm{min}$) is the power along the major (minor) ellipse axis (cf.\ inset of Fig.~\ref{fig:polEllFit}). Thus, $e = 1$ ($e = 0$) means a linear (circular) polarization state, and $e \in (0, 1)$ elliptical polarization. This interpretation is appropriate when considering a single mode such as for the wavelength-resolved polarization measurement presented in Fig.~\ref{fig:spectraPol}. % paragraph

In the case of partially polarized light, which is evidently the case when considering the total emission from a multimode laser, Eq.~(\ref{eq:polEllFit}) describes the incoherent superposition of several modes with two different, orthogonal polarization axes, where all the individual modes are either linearly or elliptically polarized along $\alpha_0$ or $\alpha_0 + \pi/2$. In this case $\alpha_0$ is the axis of the predominant polarization state, and the eccentricity $e$ is related to the ratio of the power along the two axes, with $e = 1$ meaning completely linearly polarized light along $\alpha_0$ and $e = 0$ equal power along both axes. This is the case considered in Fig.~\ref{fig:LIcurve}. % paragraph

The measurement with a rotating polarizer and the fit with Eq.~(\ref{eq:polEllFit}) are a simple, efficient and sufficient way of characterizing the polarization state of the VCSEL considered here. However, Eq.~(\ref{eq:polEllFit}) is not applicable to all situations, for example when considering superpositions of linearly or elliptically polarized modes with non-orthogonal axes \cite{Virte2013}. Additionally, the degree of polarization cannot be directly determined, and spectrally resolved measurements are thus needed for multimode lasers. However, more refined measurement and analysis techniques have been developed to obtain the complete set of Stokes parameters characterizing the polarization state of a VCSEL \cite{Molitor2012, Molitor2015, Molitor2016}. % paragraph

\section{Fluctuations of the lasing spectrum} \label{app:specFluc}

\begin{figure}[tb]
\begin{center}
\includegraphics[width = 7 cm]{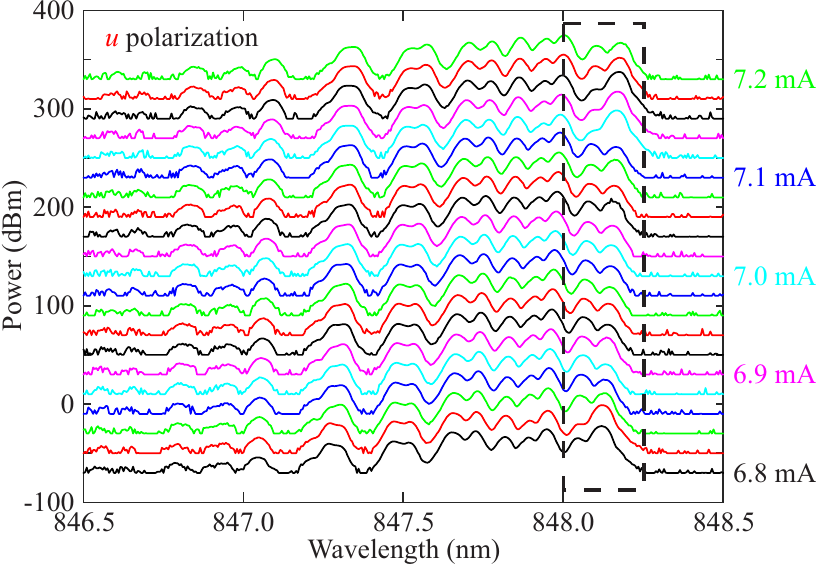}
\end{center}
\caption{Fluctuations of lasing spectra for pump currents $6.8$--$7.2$~mA with polarizer along the $u$ axis. The spectra are offset vertically by $20$~dBm for clarity. The dashed black box indicates the two lasing modes showing fluctuations of their amplitudes.}
\label{fig:spectraFluc}
\end{figure}

In some pump current regimes, the lasing spectra show erratic fluctuations when changing the current. An example is shown in Fig.~\ref{fig:spectraFluc}, where the spectra for $u$ polarization in the range of $6.8$--$7.2$~mA are shown with an increment of $0.02$~mA (same data as in Fig.~\ref{fig:spectraMainAxes}). Here, the amplitudes of the two lowest-order modes (indicated by the dashed box) change unpredictably from one current value to the next. Interestingly, the other modes in the spectrum do not exhibit such fluctuations. This indicates that which of these two modes dominates over the other depends sensitively on the system parameters, whereas the state of the other modes is less sensitive. % paragraph

One possible explanation is that the system exhibits bistability, and changes of the pump current can thus make the system switch unpredictably between these two lasing states. However, measurements with decreasing pump current showed no signs of hysteresis. Another possibility is that small fluctuations of the ambient temperature induce these fluctuations in spite of the stabilization of the diode laser mount temperature and a plastic wrap to thermally insulate the mount additionally. However, at present we cannot determine with certainty the mechanism responsible for the fluctuations. It should be noted that similar fluctuations are seen in the RF-spectra in Fig.~\ref{fig:bifurcations}(b) at $9.43$ and $9.74$~mA, and we believe they can be attributed to the same mechanism. % paragraph

\section{Evolution of near-field intensity distributions} \label{app:NFevo}

\begin{figure}[tb]
\begin{center}
\includegraphics[width = 7 cm]{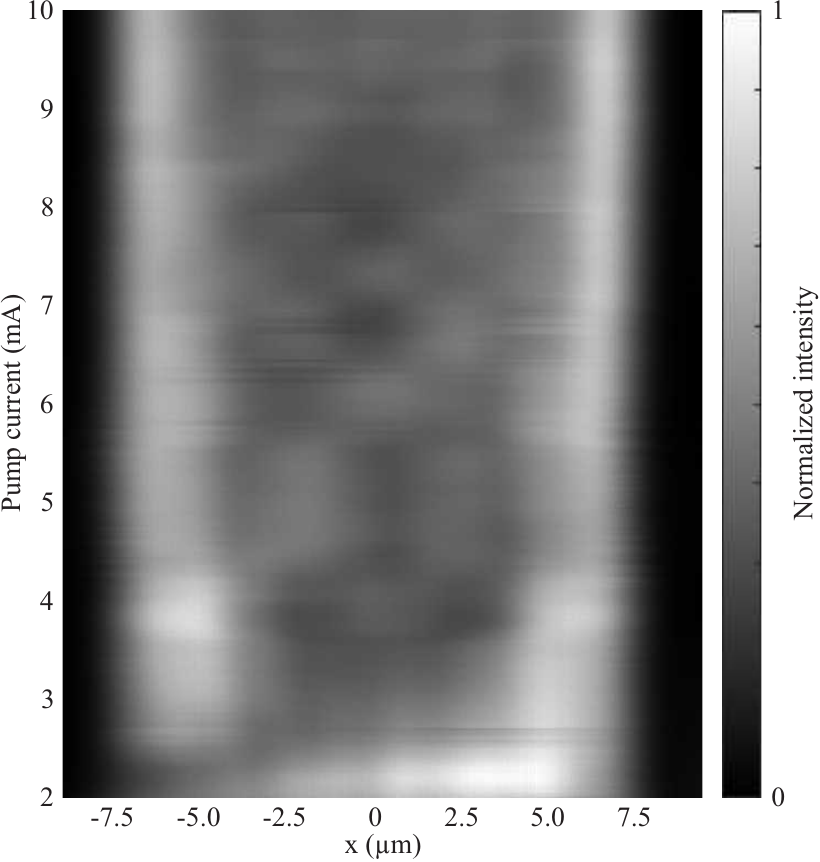}
\end{center}
\caption{Horizontal cross section of the near-field images measured without polarizer. The gray scale is separately normalized for each pump current.}
\label{fig:NFsection}
\end{figure}

Figure~\ref{fig:NFsection} shows the horizontal cross sections of the near-field (NF) images measured without polarizer (cf.\ Fig.~\ref{fig:NFimgs}) for different pump currents. While the NF intensity distributions are fairly homogeneous close to threshold, they start to be concentrated at the outer edge of the active region beyond $2.5$~mA. As the pump current is increased further, this region of high intensity becomes even narrower because transverse modes with increasing azimuthal order $m$ are excited. This is attributed to the ring-shaped top contact of the VCSEL which results in a higher current density near the outer edge of the active region \cite{Degen1999} that preferentially excites the high-order modes localized there. The intensity pattern in the center region with lower intensity also changes with the pump current as different combinations of low-order transverse modes lase. % paragraph

\section{Spatio-spectral measurements} \label{app:spatio-spec}
Spatio-spectral images of the VCSEL were measured using an imaging spectrometer (Princeton Instruments SpectraPro HRS-500) with a $1800$~g/mm grating (spectral resolution about $30$~pm) in order to identify the different transverse modes. The input slit of the spectrometer was opened to $1$~mm and aligned with the image plane of the VCSEL. A CCD camera with $2.2~\mu$m pixel size (Allied Vision Mako U503-B) in the focal plane of the spectrometer recorded the spatio-spectral images, that is, spectrally-resolved NF images. Furthermore the objective (Obj1 in Fig.~\ref{fig:setup}) was replaced by one with smaller magnification ($16\times$, NA $= 0.30$) to reduce the lateral overlap of the different modes in the focal plane of the spectrometer. The polarizer was aligned with the $u$ and $v$ polarization axes. Spatio-spectral measurements of BA-VCSELs using an imaging spectrometer \cite{ChangHasnain1991, Degen2000, Debernardi2002, Misak2015} or a state tomography technique \cite{Ploeschner2022} have been reported previously. % paragraph

\begin{figure*}[tb]
\begin{center}
\includegraphics[width = 15 cm]{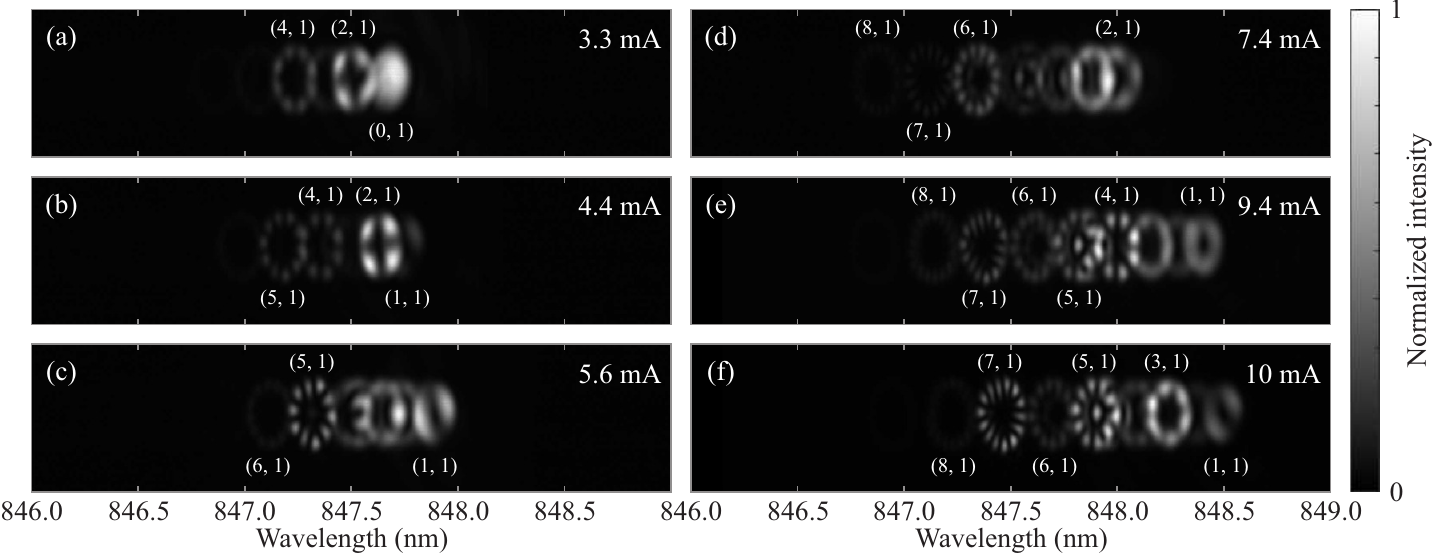}
\end{center}
\caption{Spatio-spectral images for $u$ polarization and pump currents (a)~$3.3$~mA, (b)~$4.4$~mA, (c)~$5.6$~mA, (d)~$7.4$~mA, (e)~$9.4$~mA, and (f)~$10.0$~mA. The quantum numbers $(m, n)$ are indicated for the modes that can be clearly identified.}
\label{fig:SpatioSpecID}
\end{figure*}

The spatio-spectral images for $u$ and $v$ polarization were measured systematically in the range of $2$--$10$~mA with a current step of $0.1$~mA. Several examples of the measured spatio-spectral images are presented in Fig.~\ref{fig:SpatioSpecID}. The transverse modes are labeled by their azimuthal and radial quantum numbers $(m, n)$. It should be noted that the modes with $m > 0$ are two-fold degenerate, where one mode has an azimuthal dependence of the intensity $I \propto \cos^2(m \varphi - \varphi_0)$ and the other one $I \propto \sin^2(m \varphi - \varphi_0)$. Since most of the time only either the cos- or the sin-type of a given transverse mode lases in one polarization, we can identify the modes by counting the number of maxima in azimuthal and radial direction. The indices of the clearly identifiable modes are indicated in Fig.~\ref{fig:SpatioSpecID}. Moreover, by comparing the images for different pump currents and following the evolution of the measured spectra taking into account the thermal red-shift, we can effectively identify all the lasing modes, and their quantum numbers are indicated at the top in Fig.~\ref{fig:bifurcations}(a). % paragraph

The good spectral and spatial resolution of the images furthermore allows to investigate the structure of the different transverse modes in more detail. In an ideal VCSEL with rotationally symmetric structure, the patterns of the transverse modes should also exhibit circular symmetry. However, we find that some modes exhibit significant deviations from the expected symmetry. For example, the fundamental mode $(0, 1)$ shows higher intensity towards the upper right corner, which explains the same observation in the NF images near threshold [cf.\ Fig.~\ref{fig:NFimgs}(a)]. This could be caused for example by inhomogeneities in the active layer of the VCSEL. % pargraph

Furthermore, some of the higher-order modes show a polygonal instead of circular structure for some current regimes, notably the $(5, 1)$ mode in Fig.~\ref{fig:SpatioSpecID}(c) and the $(7, 1)$ mode in Figs.~\ref{fig:SpatioSpecID}(d, e, f). In contrast, most of the other modes do not show such deformations for any pump current. While the cause of these polygonal deformations is not clear, we can exclude structural defects and thermal lensing: first, the deformed modes coexist with other high-order modes without visible deformation at the same current, whereas defects or thermal lenses should affect all modes in a similar manner. Second, the degree of deformation changes with the pump current, for example the $(5, 1)$ mode exhibits a quite circular structure in Figs.~\ref{fig:SpatioSpecID}(b, e, f). These observations indicate that dynamic effects are probably responsible. % paragraph

Another important question is what determines the azimuthal orientation of the modes (that is, $\varphi_0$). While pinning by structural defects and inhomogeneities probably plays a role \cite{Pereira1998}, there are several observations which indicate that dynamic effects are important as well. For example, the symmetry axes of the $(4, 1)$ mode in Figs.~\ref{fig:SpatioSpecID}(a, b) are well aligned with the vertical and horizontal axes. In contrast, the symmetry axes of the $(2, 1)$ mode are visibly tilted. The same applies for other modes such as the $(5, 1)$ mode in Fig.~\ref{fig:SpatioSpecID}(b). However, if defect pinning was the main factor determining the mode orientation, we would expect that the transverse modes all have the same orientation. % pargraph

\begin{figure*}[tb]
\begin{center}
\includegraphics[width = 15 cm]{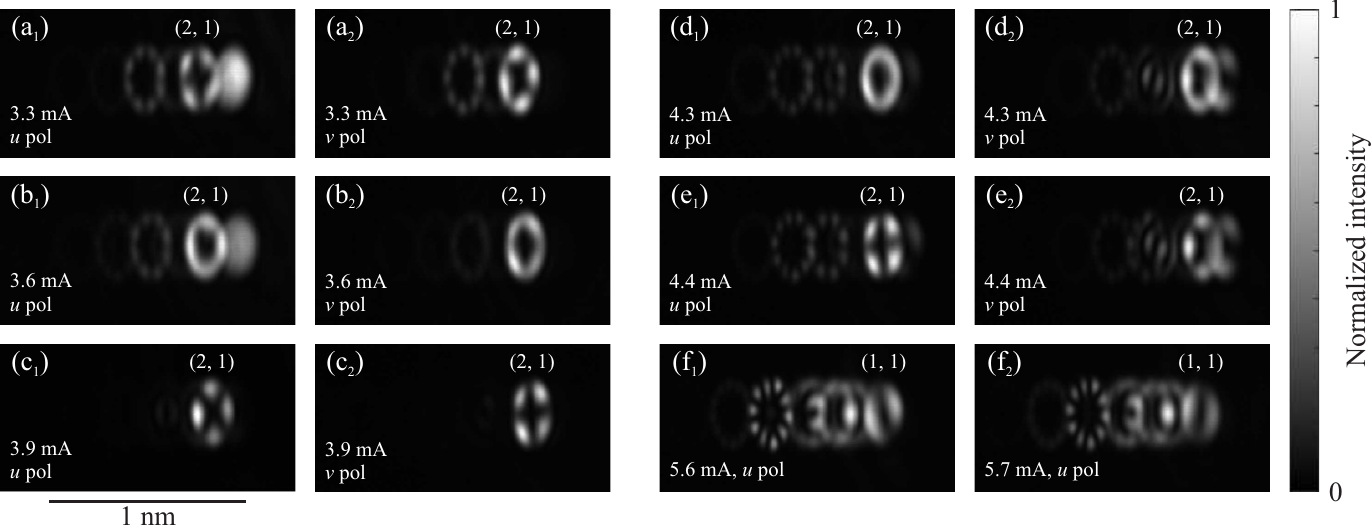}
\end{center}
\caption{Changes of azimuthal mode orientation. (a)--(e)~Spatio-spectral images showing the evolution of the $(2, 1)$ mode from $3.3$ to $4.4$~mA. The left (right) panels show the $u$ ($v$) polarization. (f)~Spatio-spectral images showing the rotation of the $(1, 1)$ mode with $u$ polarization from $5.6$ to $5.7$~mA. The horizontal wavelength axis of the images covers a spectral range of $1.5$~nm. The $(2, 1)$ and the $(1, 1)$ modes are highlighted with their quantum numbers.}
\label{fig:modeRotation}
\end{figure*}

Moreover, some modes change their orientation with increasing pump current (cf.\ Fig.~\ref{fig:imgOrient}). Two examples are shown in Fig.~\ref{fig:modeRotation}. At $3.3$~mA [Fig.~\ref{fig:modeRotation}(a)], the $(2, 1)$ mode lases in both polarizations, but with $90^\circ$-shifted orientation, that is, the cos-type mode is excited in one and the sin-type mode in the other polariation. As the pump current increases, both polarizations start to exhibit the cos- as well as the sin-type mode, so that for $3.6$~mA an almost homogeneous ring is observed in both polarizations [Fig.~\ref{fig:modeRotation}(b)]. As the current increases further to $3.9$~mA [Fig.~\ref{fig:modeRotation}(c)], the VCSEL goes back to lasing in only one of the types for each polarization. However, the orientation of the $(2, 1)$ mode was interchanged between the two polarizations. This evolution repeats itself again when increasing the current further to $4.4$~mA as shown in Figs.~\ref{fig:modeRotation}(d, e). In summary, for each polarization we see a rotation of the orientation of the $(2, 1)$ mode that passes via a superposition of cos- and sin-type modes, but the basic orientation of the modes (i.e., the value of $\varphi_0$) does not change significantly as seen in Figs.~\ref{fig:modeRotation}(a, c, e). It seems that one (probably static) mechanism fixes the orientation of the modes, whereas another, current-dependent mechanism determines which of the orientations (cos- or sin-type) is excited in which polarization. % paragraph

The example of the $(2, 1)$ mode also demonstrates again that the transverse modes try to lase in a different orientation for each polarization, which is found for the majority of cases. However, it is also possible for both orientations to lase in the same polarization, and even in both polarizations at the same time, as shown in Figs.~\ref{fig:modeRotation}(b, d). So while gain competition clearly favors the first scenario, it does not completely prohibit the latter one either. % paragraph 

A different type of rotation observed for the $(1, 1)$ mode is shown in Fig.~\ref{fig:modeRotation}(f). When increasing the pump current from $5.6$~mA [Fig.~\ref{fig:modeRotation}(f$_1$)] to $5.7$~mA [Fig.~\ref{fig:modeRotation}(f$_2$)], the orientation of the $(1, 1)$ mode with $u$ polarization rotates. However, this is not just a change between the cos- and sin-type mode [as observed for the $(2, 1)$ mode], since this would imply a rotation by $90^\circ$. Here the rotation angle is clearly different from $90^\circ$, which means that $\varphi_0$ changed with the pump current. In conclusion, there appear to be several mechanisms, both static and dynamic, that influence the azimuthal orientation of the transverse modes, and their interplay results in different behaviors for different modes and pump currents. % paragraph

\section{Output power fluctuations} \label{app:powFluc}

\begin{figure}[tb]
\begin{center}
\includegraphics[width = 7 cm]{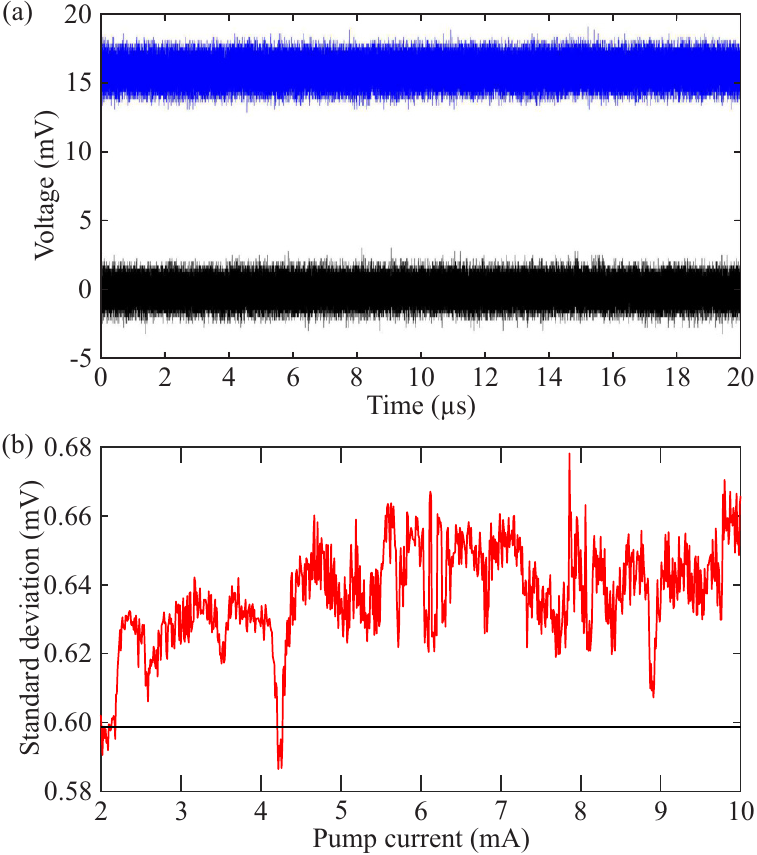}
\end{center}
\caption{(a) Time trace for $6$~mA (blue) and background signal without laser (black). Measurement with polarizer along the $v$ axis and without RF-amplifier. (b)~Standard deviation of the time traces as function of pump current (red) and of background (black).}
\label{fig:TTstd}
\end{figure}

Figure~\ref{fig:TTstd}(a) shows the time trace of the output power measured with the polarizer along the $v$ axis (blue) for $6$~mA. The DC component of the signal is preserved since no RF-amplifier is used in the measurement. The time trace appears quite stable with fluctuations of a level similar to the noise level of the measurement instruments (black). This is the case for all other pump currents and measurements without polarizer as well. Figure~\ref{fig:TTstd}(b) shows the standard deviation of the time traces as a function of the pump current (red) in comparison to that of the noise signal shown in Fig.~\ref{fig:TTstd}(a). While the fluctuations of the laser power slightly increase with pump current, they remain of the order of the noise signal fluctuations. Hence an RF-amplifier is needed to investigate the small fluctuations of the VCSEL output power around its mean value. Given the high complexity of the multimode VCSEL, it is not \textit{a priori} evident that it features only small output power fluctuations. % paragraph

Furthermore it should be noted that no polarization-mode hopping was observed, that is, sudden switches of the power between the two polarization states. Instead, the power in each polarization state stays almost constant in any operation regime. Polarization-mode hopping was observed for narrow VCSELs in particular close to polarization-switching points \cite{Willemsen1999, Ackemann2001, Olejniczak2011}. Its absence even close to polarization-switching points for the BA-VCSEL considered here indicates that the transverse mode competition is the more important factor driving its bifurcations. % paragraph


\begin{thebibliography}{73}%
\makeatletter
\providecommand \@ifxundefined [1]{%
 \@ifx{#1\undefined}
}%
\providecommand \@ifnum [1]{%
 \ifnum #1\expandafter \@firstoftwo
 \else \expandafter \@secondoftwo
 \fi
}%
\providecommand \@ifx [1]{%
 \ifx #1\expandafter \@firstoftwo
 \else \expandafter \@secondoftwo
 \fi
}%
\providecommand \natexlab [1]{#1}%
\providecommand \enquote  [1]{``#1''}%
\providecommand \bibnamefont  [1]{#1}%
\providecommand \bibfnamefont [1]{#1}%
\providecommand \citenamefont [1]{#1}%
\providecommand \href@noop [0]{\@secondoftwo}%
\providecommand \href [0]{\begingroup \@sanitize@url \@href}%
\providecommand \@href[1]{\@@startlink{#1}\@@href}%
\providecommand \@@href[1]{\endgroup#1\@@endlink}%
\providecommand \@sanitize@url [0]{\catcode `\\12\catcode `\$12\catcode
  `\&12\catcode `\#12\catcode `\^12\catcode `\_12\catcode `\%12\relax}%
\providecommand \@@startlink[1]{}%
\providecommand \@@endlink[0]{}%
\providecommand \url  [0]{\begingroup\@sanitize@url \@url }%
\providecommand \@url [1]{\endgroup\@href {#1}{\urlprefix }}%
\providecommand \urlprefix  [0]{URL }%
\providecommand \Eprint [0]{\href }%
\providecommand \doibase [0]{https://doi.org/}%
\providecommand \selectlanguage [0]{\@gobble}%
\providecommand \bibinfo  [0]{\@secondoftwo}%
\providecommand \bibfield  [0]{\@secondoftwo}%
\providecommand \translation [1]{[#1]}%
\providecommand \BibitemOpen [0]{}%
\providecommand \bibitemStop [0]{}%
\providecommand \bibitemNoStop [0]{.\EOS\space}%
\providecommand \EOS [0]{\spacefactor3000\relax}%
\providecommand \BibitemShut  [1]{\csname bibitem#1\endcsname}%
\let\auto@bib@innerbib\@empty
%</preamble>
\bibitem [{\citenamefont {Ohtsubo}(2013)}]{OhtsuboBook2013}%
  \BibitemOpen
  \bibfield  {author} {\bibinfo {author} {\bibfnamefont {J.}~\bibnamefont
  {Ohtsubo}},\ }\href {https://doi.org/10.1007/978-3-642-30147-6} {\emph
  {\bibinfo {title} {Semiconductor Lasers - Stability, Instability and
  Chaos}}},\ \bibinfo {edition} {3rd}\ ed.\ (\bibinfo  {publisher} {Springer},\
  \bibinfo {address} {Heidelberg},\ \bibinfo {year} {2013})\BibitemShut
  {NoStop}%
\bibitem [{\citenamefont {Abraham}\ \emph {et~al.}(1985)\citenamefont
  {Abraham}, \citenamefont {Lugiato},\ and\ \citenamefont
  {Narducci}}]{Abraham1985}%
  \BibitemOpen
  \bibfield  {author} {\bibinfo {author} {\bibfnamefont {N.~B.}\ \bibnamefont
  {Abraham}}, \bibinfo {author} {\bibfnamefont {L.~A.}\ \bibnamefont
  {Lugiato}},\ and\ \bibinfo {author} {\bibfnamefont {L.~M.}\ \bibnamefont
  {Narducci}},\ }\bibfield  {title} {\bibinfo {title} {Overview of
  instabilities in laser systems},\ }\href
  {https://doi.org/10.1364/JOSAB.2.000007} {\bibfield  {journal} {\bibinfo
  {journal} {J. Opt. Soc. Am. B}\ }\textbf {\bibinfo {volume} {2}},\ \bibinfo
  {pages} {7} (\bibinfo {year} {1985})}\BibitemShut {NoStop}%
\bibitem [{\citenamefont {Weiss}(1988)}]{Weiss1988}%
  \BibitemOpen
  \bibfield  {author} {\bibinfo {author} {\bibfnamefont {C.~O.}\ \bibnamefont
  {Weiss}},\ }\bibfield  {title} {\bibinfo {title} {Chaotic laser dynamics},\
  }\href {https://doi.org/10.1007/BF02069690} {\bibfield  {journal} {\bibinfo
  {journal} {Opt. Quantum Electr.}\ }\textbf {\bibinfo {volume} {20}},\
  \bibinfo {pages} {1} (\bibinfo {year} {1988})}\BibitemShut {NoStop}%
\bibitem [{\citenamefont {Sciamanna}\ and\ \citenamefont
  {Shore}(2015)}]{Sciamanna2015}%
  \BibitemOpen
  \bibfield  {author} {\bibinfo {author} {\bibfnamefont {M.}~\bibnamefont
  {Sciamanna}}\ and\ \bibinfo {author} {\bibfnamefont {K.~A.}\ \bibnamefont
  {Shore}},\ }\bibfield  {title} {\bibinfo {title} {Physics and applications of
  laser diode chaos},\ }\href {https://doi.org/10.1038/nphoton.2014.326}
  {\bibfield  {journal} {\bibinfo  {journal} {Nature Photonics}\ }\textbf
  {\bibinfo {volume} {9}},\ \bibinfo {pages} {151} (\bibinfo {year}
  {2015})}\BibitemShut {NoStop}%
\bibitem [{\citenamefont {Uchida}\ \emph {et~al.}(2008)\citenamefont {Uchida},
  \citenamefont {Amano}, \citenamefont {Inoue}, \citenamefont {Hirano},
  \citenamefont {Naito}, \citenamefont {Someya}, \citenamefont {Oowada},
  \citenamefont {Kurashige}, \citenamefont {Shiki}, \citenamefont {Yoshimori},
  \citenamefont {Yoshimura},\ and\ \citenamefont {Davis}}]{Uchida2008}%
  \BibitemOpen
  \bibfield  {author} {\bibinfo {author} {\bibfnamefont {A.}~\bibnamefont
  {Uchida}}, \bibinfo {author} {\bibfnamefont {K.}~\bibnamefont {Amano}},
  \bibinfo {author} {\bibfnamefont {M.}~\bibnamefont {Inoue}}, \bibinfo
  {author} {\bibfnamefont {K.}~\bibnamefont {Hirano}}, \bibinfo {author}
  {\bibfnamefont {S.}~\bibnamefont {Naito}}, \bibinfo {author} {\bibfnamefont
  {H.}~\bibnamefont {Someya}}, \bibinfo {author} {\bibfnamefont
  {I.}~\bibnamefont {Oowada}}, \bibinfo {author} {\bibfnamefont
  {T.}~\bibnamefont {Kurashige}}, \bibinfo {author} {\bibfnamefont
  {M.}~\bibnamefont {Shiki}}, \bibinfo {author} {\bibfnamefont
  {S.}~\bibnamefont {Yoshimori}}, \bibinfo {author} {\bibfnamefont
  {K.}~\bibnamefont {Yoshimura}},\ and\ \bibinfo {author} {\bibfnamefont
  {P.}~\bibnamefont {Davis}},\ }\bibfield  {title} {\bibinfo {title} {Fast
  physical random bit generation with chaotic semiconductor lasers},\ }\href
  {https://doi.org/10.1038/nphoton.2008.227} {\bibfield  {journal} {\bibinfo
  {journal} {Nature Photonics}\ }\textbf {\bibinfo {volume} {2}},\ \bibinfo
  {pages} {728} (\bibinfo {year} {2008})}\BibitemShut {NoStop}%
\bibitem [{\citenamefont {Qi}\ and\ \citenamefont {Liu}(2011)}]{Qi2011}%
  \BibitemOpen
  \bibfield  {author} {\bibinfo {author} {\bibfnamefont {X.-Q.}\ \bibnamefont
  {Qi}}\ and\ \bibinfo {author} {\bibfnamefont {J.-M.}\ \bibnamefont {Liu}},\
  }\bibfield  {title} {\bibinfo {title} {Photonic microwave applications of the
  dynamics of semiconductor lasers},\ }\href
  {https://doi.org/10.1109/JSTQE.2011.2121055} {\bibfield  {journal} {\bibinfo
  {journal} {IEEE J. Sel. Top. Quant. Electron.}\ }\textbf {\bibinfo {volume}
  {17}},\ \bibinfo {pages} {1198} (\bibinfo {year} {2011})}\BibitemShut
  {NoStop}%
\bibitem [{\citenamefont {Lenstra}\ \emph {et~al.}(1985)\citenamefont
  {Lenstra}, \citenamefont {Verbeek},\ and\ \citenamefont
  {Den~Boef}}]{Lenstra1985}%
  \BibitemOpen
  \bibfield  {author} {\bibinfo {author} {\bibfnamefont {D.}~\bibnamefont
  {Lenstra}}, \bibinfo {author} {\bibfnamefont {B.}~\bibnamefont {Verbeek}},\
  and\ \bibinfo {author} {\bibfnamefont {A.}~\bibnamefont {Den~Boef}},\
  }\bibfield  {title} {\bibinfo {title} {Coherence collapse in single-mode
  semiconductor lasers due to optical feedback},\ }\href
  {https://doi.org/10.1109/JQE.1985.1072725} {\bibfield  {journal} {\bibinfo
  {journal} {IEEE J. Quant. Electron.}\ }\textbf {\bibinfo {volume} {21}},\
  \bibinfo {pages} {674} (\bibinfo {year} {1985})}\BibitemShut {NoStop}%
\bibitem [{\citenamefont {Soriano}\ \emph {et~al.}(2013)\citenamefont
  {Soriano}, \citenamefont {Garc\'{\i}a-Ojalvo}, \citenamefont {Mirasso},\ and\
  \citenamefont {Fischer}}]{Soriano2013}%
  \BibitemOpen
  \bibfield  {author} {\bibinfo {author} {\bibfnamefont {M.~C.}\ \bibnamefont
  {Soriano}}, \bibinfo {author} {\bibfnamefont {J.}~\bibnamefont
  {Garc\'{\i}a-Ojalvo}}, \bibinfo {author} {\bibfnamefont {C.~R.}\ \bibnamefont
  {Mirasso}},\ and\ \bibinfo {author} {\bibfnamefont {I.}~\bibnamefont
  {Fischer}},\ }\bibfield  {title} {\bibinfo {title} {Complex photonics:
  Dynamics and applications of delay-coupled semiconductors lasers},\ }\href
  {https://doi.org/10.1103/RevModPhys.85.421} {\bibfield  {journal} {\bibinfo
  {journal} {Rev. Mod. Phys.}\ }\textbf {\bibinfo {volume} {85}},\ \bibinfo
  {pages} {421} (\bibinfo {year} {2013})}\BibitemShut {NoStop}%
\bibitem [{\citenamefont {Rontani}\ \emph {et~al.}(2007)\citenamefont
  {Rontani}, \citenamefont {Locquet}, \citenamefont {Sciamanna},\ and\
  \citenamefont {Citrin}}]{Rontani2007}%
  \BibitemOpen
  \bibfield  {author} {\bibinfo {author} {\bibfnamefont {D.}~\bibnamefont
  {Rontani}}, \bibinfo {author} {\bibfnamefont {A.}~\bibnamefont {Locquet}},
  \bibinfo {author} {\bibfnamefont {M.}~\bibnamefont {Sciamanna}},\ and\
  \bibinfo {author} {\bibfnamefont {D.~S.}\ \bibnamefont {Citrin}},\ }\bibfield
   {title} {\bibinfo {title} {Loss of time-delay signature in the chaotic
  output of a semiconductor laser with optical feedback},\ }\href
  {https://doi.org/10.1364/OL.32.002960} {\bibfield  {journal} {\bibinfo
  {journal} {Opt. Lett.}\ }\textbf {\bibinfo {volume} {32}},\ \bibinfo {pages}
  {2960} (\bibinfo {year} {2007})}\BibitemShut {NoStop}%
\bibitem [{\citenamefont {Paoli}\ and\ \citenamefont
  {Ripper}(1969)}]{Paoli1969}%
  \BibitemOpen
  \bibfield  {author} {\bibinfo {author} {\bibfnamefont {T.~L.}\ \bibnamefont
  {Paoli}}\ and\ \bibinfo {author} {\bibfnamefont {J.~E.}\ \bibnamefont
  {Ripper}},\ }\bibfield  {title} {\bibinfo {title} {Coupled longitudinal mode
  pulsing in semiconductor lasers},\ }\href
  {https://doi.org/10.1103/PhysRevLett.22.1085} {\bibfield  {journal} {\bibinfo
   {journal} {Phys. Rev. Lett.}\ }\textbf {\bibinfo {volume} {22}},\ \bibinfo
  {pages} {1085} (\bibinfo {year} {1969})}\BibitemShut {NoStop}%
\bibitem [{\citenamefont {Furfaro}\ \emph {et~al.}(2004)\citenamefont
  {Furfaro}, \citenamefont {Pedaci}, \citenamefont {Giudici}, \citenamefont
  {Hachair}, \citenamefont {Tredicce},\ and\ \citenamefont
  {Balac}}]{Furfaro2004}%
  \BibitemOpen
  \bibfield  {author} {\bibinfo {author} {\bibfnamefont {L.}~\bibnamefont
  {Furfaro}}, \bibinfo {author} {\bibfnamefont {F.}~\bibnamefont {Pedaci}},
  \bibinfo {author} {\bibfnamefont {M.}~\bibnamefont {Giudici}}, \bibinfo
  {author} {\bibfnamefont {X.}~\bibnamefont {Hachair}}, \bibinfo {author}
  {\bibfnamefont {J.}~\bibnamefont {Tredicce}},\ and\ \bibinfo {author}
  {\bibfnamefont {S.}~\bibnamefont {Balac}},\ }\bibfield  {title} {\bibinfo
  {title} {Mode-switching in semiconductor lasers},\ }\href
  {https://doi.org/10.1109/JQE.2004.834772} {\bibfield  {journal} {\bibinfo
  {journal} {IEEE J. Quantum Electron.}\ }\textbf {\bibinfo {volume} {40}},\
  \bibinfo {pages} {1365} (\bibinfo {year} {2004})}\BibitemShut {NoStop}%
\bibitem [{\citenamefont {Virte}\ \emph
  {et~al.}(2013{\natexlab{a}})\citenamefont {Virte}, \citenamefont {Panajotov},
  \citenamefont {Thienpont},\ and\ \citenamefont {Sciamanna}}]{Virte2013}%
  \BibitemOpen
  \bibfield  {author} {\bibinfo {author} {\bibfnamefont {M.}~\bibnamefont
  {Virte}}, \bibinfo {author} {\bibfnamefont {K.}~\bibnamefont {Panajotov}},
  \bibinfo {author} {\bibfnamefont {H.}~\bibnamefont {Thienpont}},\ and\
  \bibinfo {author} {\bibfnamefont {M.}~\bibnamefont {Sciamanna}},\ }\bibfield
  {title} {\bibinfo {title} {Deterministic polarization chaos from a laser
  diode},\ }\href {https://doi.org/10.1038/nphoton.2012.286} {\bibfield
  {journal} {\bibinfo  {journal} {Nature Photonics}\ }\textbf {\bibinfo
  {volume} {7}},\ \bibinfo {pages} {60} (\bibinfo {year}
  {2013}{\natexlab{a}})}\BibitemShut {NoStop}%
\bibitem [{\citenamefont {Fischer}\ \emph {et~al.}(1996)\citenamefont
  {Fischer}, \citenamefont {Hess}, \citenamefont {Els\"a{\ss}er},\ and\
  \citenamefont {G\"obel}}]{Fischer1996}%
  \BibitemOpen
  \bibfield  {author} {\bibinfo {author} {\bibfnamefont {I.}~\bibnamefont
  {Fischer}}, \bibinfo {author} {\bibfnamefont {O.}~\bibnamefont {Hess}},
  \bibinfo {author} {\bibfnamefont {W.}~\bibnamefont {Els\"a{\ss}er}},\ and\
  \bibinfo {author} {\bibfnamefont {E.}~\bibnamefont {G\"obel}},\ }\bibfield
  {title} {\bibinfo {title} {Complex spatio-temporal dynamics in the near-field
  of a broad-area semiconductor laser},\ }\href
  {https://doi.org/10.1209/epl/i1996-00154-7} {\bibfield  {journal} {\bibinfo
  {journal} {Europhys. Lett.}\ }\textbf {\bibinfo {volume} {35}},\ \bibinfo
  {pages} {579} (\bibinfo {year} {1996})}\BibitemShut {NoStop}%
\bibitem [{\citenamefont {Marciante}\ and\ \citenamefont
  {Agrawal}(1998)}]{Marciante1998}%
  \BibitemOpen
  \bibfield  {author} {\bibinfo {author} {\bibfnamefont {J.~R.}\ \bibnamefont
  {Marciante}}\ and\ \bibinfo {author} {\bibfnamefont {G.~P.}\ \bibnamefont
  {Agrawal}},\ }\bibfield  {title} {\bibinfo {title} {Spatio-temporal
  characteristics of filamentation in broad-area semiconductor lasers:
  experimental results},\ }\href {https://doi.org/10.1109/68.651101} {\bibfield
   {journal} {\bibinfo  {journal} {IEEE Photon. Technol. Lett.}\ }\textbf
  {\bibinfo {volume} {10}},\ \bibinfo {pages} {54} (\bibinfo {year}
  {1998})}\BibitemShut {NoStop}%
\bibitem [{\citenamefont {Scholz}\ \emph {et~al.}(2008)\citenamefont {Scholz},
  \citenamefont {Braun}, \citenamefont {Schwarz}, \citenamefont
  {Br\"{u}ninghoff}, \citenamefont {Queren}, \citenamefont {Lell},\ and\
  \citenamefont {Strauss}}]{Scholz2008}%
  \BibitemOpen
  \bibfield  {author} {\bibinfo {author} {\bibfnamefont {D.}~\bibnamefont
  {Scholz}}, \bibinfo {author} {\bibfnamefont {H.}~\bibnamefont {Braun}},
  \bibinfo {author} {\bibfnamefont {U.~T.}\ \bibnamefont {Schwarz}}, \bibinfo
  {author} {\bibfnamefont {S.}~\bibnamefont {Br\"{u}ninghoff}}, \bibinfo
  {author} {\bibfnamefont {D.}~\bibnamefont {Queren}}, \bibinfo {author}
  {\bibfnamefont {A.}~\bibnamefont {Lell}},\ and\ \bibinfo {author}
  {\bibfnamefont {U.}~\bibnamefont {Strauss}},\ }\bibfield  {title} {\bibinfo
  {title} {Measurement and simulation of filamentation in (al,in)gan laser
  diodes},\ }\href {https://doi.org/10.1364/OE.16.006846} {\bibfield  {journal}
  {\bibinfo  {journal} {Opt. Express}\ }\textbf {\bibinfo {volume} {16}},\
  \bibinfo {pages} {6846} (\bibinfo {year} {2008})}\BibitemShut {NoStop}%
\bibitem [{\citenamefont {Arahata}\ and\ \citenamefont
  {Uchida}(2015)}]{Arahata2015}%
  \BibitemOpen
  \bibfield  {author} {\bibinfo {author} {\bibfnamefont {M.}~\bibnamefont
  {Arahata}}\ and\ \bibinfo {author} {\bibfnamefont {A.}~\bibnamefont
  {Uchida}},\ }\bibfield  {title} {\bibinfo {title} {Inphase and antiphase
  dynamics of spatially-resolved light intensities emitted by a chaotic
  broad-area semiconductor laser},\ }\href
  {https://doi.org/10.1109/JSTQE.2015.2422473} {\bibfield  {journal} {\bibinfo
  {journal} {IEEE J. Sel. Top. Quant. Electron.}\ }\textbf {\bibinfo {volume}
  {21}},\ \bibinfo {pages} {1800609} (\bibinfo {year} {2015})}\BibitemShut
  {NoStop}%
\bibitem [{\citenamefont {Ma}\ \emph {et~al.}(2022)\citenamefont {Ma},
  \citenamefont {Xiao}, \citenamefont {Xiao}, \citenamefont {Yang},\ and\
  \citenamefont {Huang}}]{Ma2022a}%
  \BibitemOpen
  \bibfield  {author} {\bibinfo {author} {\bibfnamefont {C.-G.}\ \bibnamefont
  {Ma}}, \bibinfo {author} {\bibfnamefont {J.-L.}\ \bibnamefont {Xiao}},
  \bibinfo {author} {\bibfnamefont {Z.-X.}\ \bibnamefont {Xiao}}, \bibinfo
  {author} {\bibfnamefont {Y.-D.}\ \bibnamefont {Yang}},\ and\ \bibinfo
  {author} {\bibfnamefont {Y.-Z.}\ \bibnamefont {Huang}},\ }\bibfield  {title}
  {\bibinfo {title} {Chaotic microlasers caused by internal mode interaction
  for random number generation},\ }\href
  {https://doi.org/10.1038/s41377-022-00890-w} {\bibfield  {journal} {\bibinfo
  {journal} {Light: Sci. Appl.}\ }\textbf {\bibinfo {volume} {11}},\ \bibinfo
  {pages} {187} (\bibinfo {year} {2022})}\BibitemShut {NoStop}%
\bibitem [{\citenamefont {Bittner}\ \emph {et~al.}(2018)\citenamefont
  {Bittner}, \citenamefont {Guazzotti}, \citenamefont {Zeng}, \citenamefont
  {Hu}, \citenamefont {Y{\i}lmaz}, \citenamefont {Kim}, \citenamefont {Oh},
  \citenamefont {Wang}, \citenamefont {Hess},\ and\ \citenamefont
  {Cao}}]{Bittner2018a}%
  \BibitemOpen
  \bibfield  {author} {\bibinfo {author} {\bibfnamefont {S.}~\bibnamefont
  {Bittner}}, \bibinfo {author} {\bibfnamefont {S.}~\bibnamefont {Guazzotti}},
  \bibinfo {author} {\bibfnamefont {Y.}~\bibnamefont {Zeng}}, \bibinfo {author}
  {\bibfnamefont {X.}~\bibnamefont {Hu}}, \bibinfo {author} {\bibfnamefont
  {H.}~\bibnamefont {Y{\i}lmaz}}, \bibinfo {author} {\bibfnamefont
  {K.}~\bibnamefont {Kim}}, \bibinfo {author} {\bibfnamefont {S.~S.}\
  \bibnamefont {Oh}}, \bibinfo {author} {\bibfnamefont {Q.~J.}\ \bibnamefont
  {Wang}}, \bibinfo {author} {\bibfnamefont {O.}~\bibnamefont {Hess}},\ and\
  \bibinfo {author} {\bibfnamefont {H.}~\bibnamefont {Cao}},\ }\bibfield
  {title} {\bibinfo {title} {Suppressing spatio-temporal lasing instabilities
  with wave-chaotic microcavities},\ }\href
  {https://doi.org/10.1126/science.aas9437} {\bibfield  {journal} {\bibinfo
  {journal} {Science}\ }\textbf {\bibinfo {volume} {361}},\ \bibinfo {pages}
  {1225} (\bibinfo {year} {2018})}\BibitemShut {NoStop}%
\bibitem [{\citenamefont {Kim}\ \emph {et~al.}(2022)\citenamefont {Kim},
  \citenamefont {Bittner}, \citenamefont {Jin}, \citenamefont {Zeng},
  \citenamefont {Guazzotti}, \citenamefont {Hess}, \citenamefont {Wang},\ and\
  \citenamefont {Cao}}]{Kim2022}%
  \BibitemOpen
  \bibfield  {author} {\bibinfo {author} {\bibfnamefont {K.}~\bibnamefont
  {Kim}}, \bibinfo {author} {\bibfnamefont {S.}~\bibnamefont {Bittner}},
  \bibinfo {author} {\bibfnamefont {Y.}~\bibnamefont {Jin}}, \bibinfo {author}
  {\bibfnamefont {Y.}~\bibnamefont {Zeng}}, \bibinfo {author} {\bibfnamefont
  {S.}~\bibnamefont {Guazzotti}}, \bibinfo {author} {\bibfnamefont
  {O.}~\bibnamefont {Hess}}, \bibinfo {author} {\bibfnamefont {Q.~J.}\
  \bibnamefont {Wang}},\ and\ \bibinfo {author} {\bibfnamefont
  {H.}~\bibnamefont {Cao}},\ }\bibfield  {title} {\bibinfo {title} {Sensitive
  control of broad-area semiconductor lasers by cavity shape},\ }\href
  {https://doi.org/10.1063/5.0087048} {\bibfield  {journal} {\bibinfo
  {journal} {APL Phot.}\ }\textbf {\bibinfo {volume} {7}},\ \bibinfo {pages}
  {056106} (\bibinfo {year} {2022})}\BibitemShut {NoStop}%
\bibitem [{\citenamefont {Kim}\ \emph {et~al.}(2019)\citenamefont {Kim},
  \citenamefont {Bittner}, \citenamefont {Zeng}, \citenamefont {Liew},
  \citenamefont {Wang},\ and\ \citenamefont {Cao}}]{Kim2019}%
  \BibitemOpen
  \bibfield  {author} {\bibinfo {author} {\bibfnamefont {K.}~\bibnamefont
  {Kim}}, \bibinfo {author} {\bibfnamefont {S.}~\bibnamefont {Bittner}},
  \bibinfo {author} {\bibfnamefont {Y.}~\bibnamefont {Zeng}}, \bibinfo {author}
  {\bibfnamefont {S.~F.}\ \bibnamefont {Liew}}, \bibinfo {author}
  {\bibfnamefont {Q.}~\bibnamefont {Wang}},\ and\ \bibinfo {author}
  {\bibfnamefont {H.}~\bibnamefont {Cao}},\ }\bibfield  {title} {\bibinfo
  {title} {Electrically pumped semiconductor laser with low spatial coherence
  and directional emission},\ }\href {https://doi.org/10.1063/1.5109234}
  {\bibfield  {journal} {\bibinfo  {journal} {Appl. Phys. Lett.}\ }\textbf
  {\bibinfo {volume} {115}},\ \bibinfo {pages} {071101} (\bibinfo {year}
  {2019})}\BibitemShut {NoStop}%
\bibitem [{\citenamefont {Cao}\ \emph {et~al.}(2019)\citenamefont {Cao},
  \citenamefont {Chriki}, \citenamefont {Bittner}, \citenamefont {Friesem},\
  and\ \citenamefont {Davidson}}]{Cao2019}%
  \BibitemOpen
  \bibfield  {author} {\bibinfo {author} {\bibfnamefont {H.}~\bibnamefont
  {Cao}}, \bibinfo {author} {\bibfnamefont {R.}~\bibnamefont {Chriki}},
  \bibinfo {author} {\bibfnamefont {S.}~\bibnamefont {Bittner}}, \bibinfo
  {author} {\bibfnamefont {A.~A.}\ \bibnamefont {Friesem}},\ and\ \bibinfo
  {author} {\bibfnamefont {N.}~\bibnamefont {Davidson}},\ }\bibfield  {title}
  {\bibinfo {title} {Complex lasers with controllable coherence},\ }\href
  {https://doi.org/10.1038/s42254-018-0010-6} {\bibfield  {journal} {\bibinfo
  {journal} {Nature Reviews Physics}\ }\textbf {\bibinfo {volume} {1}},\
  \bibinfo {pages} {156} (\bibinfo {year} {2019})}\BibitemShut {NoStop}%
\bibitem [{\citenamefont {Kim}\ \emph {et~al.}(2021)\citenamefont {Kim},
  \citenamefont {Bittner}, \citenamefont {Zeng}, \citenamefont {Guazzotti},
  \citenamefont {Hess}, \citenamefont {Wang},\ and\ \citenamefont
  {Cao}}]{Kim2021}%
  \BibitemOpen
  \bibfield  {author} {\bibinfo {author} {\bibfnamefont {K.}~\bibnamefont
  {Kim}}, \bibinfo {author} {\bibfnamefont {S.}~\bibnamefont {Bittner}},
  \bibinfo {author} {\bibfnamefont {Y.}~\bibnamefont {Zeng}}, \bibinfo {author}
  {\bibfnamefont {S.}~\bibnamefont {Guazzotti}}, \bibinfo {author}
  {\bibfnamefont {O.}~\bibnamefont {Hess}}, \bibinfo {author} {\bibfnamefont
  {Q.~J.}\ \bibnamefont {Wang}},\ and\ \bibinfo {author} {\bibfnamefont
  {H.}~\bibnamefont {Cao}},\ }\bibfield  {title} {\bibinfo {title} {Massively
  parallel ultrafast random bit generation with a chip-scale laser},\ }\href
  {https://doi.org/10.1126/science.abc2666} {\bibfield  {journal} {\bibinfo
  {journal} {Science}\ }\textbf {\bibinfo {volume} {371}},\ \bibinfo {pages}
  {948} (\bibinfo {year} {2021})}\BibitemShut {NoStop}%
\bibitem [{\citenamefont {Willemsen}\ \emph {et~al.}(1999)\citenamefont
  {Willemsen}, \citenamefont {Khalid}, \citenamefont {van Exter},\ and\
  \citenamefont {Woerdman}}]{Willemsen1999}%
  \BibitemOpen
  \bibfield  {author} {\bibinfo {author} {\bibfnamefont {M.~B.}\ \bibnamefont
  {Willemsen}}, \bibinfo {author} {\bibfnamefont {M.~U.~F.}\ \bibnamefont
  {Khalid}}, \bibinfo {author} {\bibfnamefont {M.~P.}\ \bibnamefont {van
  Exter}},\ and\ \bibinfo {author} {\bibfnamefont {J.~P.}\ \bibnamefont
  {Woerdman}},\ }\bibfield  {title} {\bibinfo {title} {Polarization switching
  of a vertical-cavity semiconductor laser as a kramers hopping problem},\
  }\href {https://doi.org/10.1103/PhysRevLett.82.4815} {\bibfield  {journal}
  {\bibinfo  {journal} {Phys. Rev. Lett.}\ }\textbf {\bibinfo {volume} {82}},\
  \bibinfo {pages} {4815} (\bibinfo {year} {1999})}\BibitemShut {NoStop}%
\bibitem [{\citenamefont {Ackemann}\ and\ \citenamefont
  {Sondermann}(2001)}]{Ackemann2001}%
  \BibitemOpen
  \bibfield  {author} {\bibinfo {author} {\bibfnamefont {T.}~\bibnamefont
  {Ackemann}}\ and\ \bibinfo {author} {\bibfnamefont {M.}~\bibnamefont
  {Sondermann}},\ }\bibfield  {title} {\bibinfo {title} {Characteristics of
  polarization switching from the low to the high frequency mode in
  vertical-cavity surface-emitting lasers},\ }\href
  {https://doi.org/10.1063/1.1375833} {\bibfield  {journal} {\bibinfo
  {journal} {Appl. Phys. Lett.}\ }\textbf {\bibinfo {volume} {78}},\ \bibinfo
  {pages} {3574} (\bibinfo {year} {2001})}\BibitemShut {NoStop}%
\bibitem [{\citenamefont {Panajotov}\ \emph {et~al.}(2008)\citenamefont
  {Panajotov}, \citenamefont {Sciamanna}, \citenamefont {Gatare}, \citenamefont
  {Arteaga},\ and\ \citenamefont {Thienpont}}]{Panajotov2008}%
  \BibitemOpen
  \bibfield  {author} {\bibinfo {author} {\bibfnamefont {K.}~\bibnamefont
  {Panajotov}}, \bibinfo {author} {\bibfnamefont {M.}~\bibnamefont
  {Sciamanna}}, \bibinfo {author} {\bibfnamefont {I.}~\bibnamefont {Gatare}},
  \bibinfo {author} {\bibfnamefont {M.~A.}\ \bibnamefont {Arteaga}},\ and\
  \bibinfo {author} {\bibfnamefont {H.}~\bibnamefont {Thienpont}},\ }\bibfield
  {title} {\bibinfo {title} {Light polarization fingerprints on nonlinear
  dynamics of vertical-cavity surface-emitting lasers},\ }\href
  {https://doi.org/10.2478/s11772-008-0033-0} {\bibfield  {journal} {\bibinfo
  {journal} {Opto-Electronics Review}\ }\textbf {\bibinfo {volume} {16}},\
  \bibinfo {pages} {337} (\bibinfo {year} {2008})}\BibitemShut {NoStop}%
\bibitem [{\citenamefont {Olejniczak}\ \emph {et~al.}(2011)\citenamefont
  {Olejniczak}, \citenamefont {Panajotov}, \citenamefont {Thienpont},
  \citenamefont {Sciamanna}, \citenamefont {Mutig}, \citenamefont {Hopfer},\
  and\ \citenamefont {Bimberg}}]{Olejniczak2011}%
  \BibitemOpen
  \bibfield  {author} {\bibinfo {author} {\bibfnamefont {L.}~\bibnamefont
  {Olejniczak}}, \bibinfo {author} {\bibfnamefont {K.}~\bibnamefont
  {Panajotov}}, \bibinfo {author} {\bibfnamefont {H.}~\bibnamefont
  {Thienpont}}, \bibinfo {author} {\bibfnamefont {M.}~\bibnamefont
  {Sciamanna}}, \bibinfo {author} {\bibfnamefont {A.}~\bibnamefont {Mutig}},
  \bibinfo {author} {\bibfnamefont {F.}~\bibnamefont {Hopfer}},\ and\ \bibinfo
  {author} {\bibfnamefont {D.}~\bibnamefont {Bimberg}},\ }\bibfield  {title}
  {\bibinfo {title} {Polarization switching and polarization mode hopping in
  quantum dot vertical-cavity surface-emitting lasers},\ }\href
  {https://doi.org/10.1364/OE.19.002476} {\bibfield  {journal} {\bibinfo
  {journal} {Opt. Express}\ }\textbf {\bibinfo {volume} {19}},\ \bibinfo
  {pages} {2476} (\bibinfo {year} {2011})}\BibitemShut {NoStop}%
\bibitem [{\citenamefont {Virte}\ \emph {et~al.}(2015)\citenamefont {Virte},
  \citenamefont {Mirisola}, \citenamefont {Sciamanna},\ and\ \citenamefont
  {Panajotov}}]{Virte2015}%
  \BibitemOpen
  \bibfield  {author} {\bibinfo {author} {\bibfnamefont {M.}~\bibnamefont
  {Virte}}, \bibinfo {author} {\bibfnamefont {E.}~\bibnamefont {Mirisola}},
  \bibinfo {author} {\bibfnamefont {M.}~\bibnamefont {Sciamanna}},\ and\
  \bibinfo {author} {\bibfnamefont {K.}~\bibnamefont {Panajotov}},\ }\bibfield
  {title} {\bibinfo {title} {Asymmetric dwell-time statistics of polarization
  chaos from free-running vcsel},\ }\href
  {https://doi.org/10.1364/OL.40.001865} {\bibfield  {journal} {\bibinfo
  {journal} {Opt. Lett.}\ }\textbf {\bibinfo {volume} {40}},\ \bibinfo {pages}
  {1865} (\bibinfo {year} {2015})}\BibitemShut {NoStop}%
\bibitem [{\citenamefont {Mart\'{i}n-Regalado}\ \emph
  {et~al.}(1996)\citenamefont {Mart\'{i}n-Regalado}, \citenamefont {Miguel},
  \citenamefont {Abraham},\ and\ \citenamefont {Prati}}]{MartinRegalado1996a}%
  \BibitemOpen
  \bibfield  {author} {\bibinfo {author} {\bibfnamefont {J.}~\bibnamefont
  {Mart\'{i}n-Regalado}}, \bibinfo {author} {\bibfnamefont {M.~S.}\
  \bibnamefont {Miguel}}, \bibinfo {author} {\bibfnamefont {N.~B.}\
  \bibnamefont {Abraham}},\ and\ \bibinfo {author} {\bibfnamefont
  {F.}~\bibnamefont {Prati}},\ }\bibfield  {title} {\bibinfo {title}
  {Polarization switching in quantum-well vertical-cavity surface-emitting
  lasers},\ }\href {https://doi.org/10.1364/OL.21.000351} {\bibfield  {journal}
  {\bibinfo  {journal} {Opt. Lett.}\ }\textbf {\bibinfo {volume} {21}},\
  \bibinfo {pages} {351} (\bibinfo {year} {1996})}\BibitemShut {NoStop}%
\bibitem [{\citenamefont {Martin-Regalado}\ \emph {et~al.}(1997)\citenamefont
  {Martin-Regalado}, \citenamefont {Prati}, \citenamefont {San~Miguel},\ and\
  \citenamefont {Abraham}}]{MartinRegalado1997b}%
  \BibitemOpen
  \bibfield  {author} {\bibinfo {author} {\bibfnamefont {J.}~\bibnamefont
  {Martin-Regalado}}, \bibinfo {author} {\bibfnamefont {F.}~\bibnamefont
  {Prati}}, \bibinfo {author} {\bibfnamefont {M.}~\bibnamefont {San~Miguel}},\
  and\ \bibinfo {author} {\bibfnamefont {N.}~\bibnamefont {Abraham}},\
  }\bibfield  {title} {\bibinfo {title} {Polarization properties of
  vertical-cavity surface-emitting lasers},\ }\href
  {https://doi.org/10.1109/3.572151} {\bibfield  {journal} {\bibinfo  {journal}
  {IEEE J. Quant. Electron.}\ }\textbf {\bibinfo {volume} {33}},\ \bibinfo
  {pages} {765} (\bibinfo {year} {1997})}\BibitemShut {NoStop}%
\bibitem [{\citenamefont {Travagnin}\ \emph {et~al.}(1996)\citenamefont
  {Travagnin}, \citenamefont {van Exter}, \citenamefont {Jansen~van Doorn},\
  and\ \citenamefont {Woerdman}}]{Travagnin1996}%
  \BibitemOpen
  \bibfield  {author} {\bibinfo {author} {\bibfnamefont {M.}~\bibnamefont
  {Travagnin}}, \bibinfo {author} {\bibfnamefont {M.~P.}\ \bibnamefont {van
  Exter}}, \bibinfo {author} {\bibfnamefont {A.~K.}\ \bibnamefont {Jansen~van
  Doorn}},\ and\ \bibinfo {author} {\bibfnamefont {J.~P.}\ \bibnamefont
  {Woerdman}},\ }\bibfield  {title} {\bibinfo {title} {Role of optical
  anisotropies in the polarization properties of surface-emitting semiconductor
  lasers},\ }\href {https://doi.org/10.1103/PhysRevA.54.1647} {\bibfield
  {journal} {\bibinfo  {journal} {Phys. Rev. A}\ }\textbf {\bibinfo {volume}
  {54}},\ \bibinfo {pages} {1647} (\bibinfo {year} {1996})}\BibitemShut
  {NoStop}%
\bibitem [{\citenamefont {{van Exter}}\ \emph
  {et~al.}(1998{\natexlab{a}})\citenamefont {{van Exter}}, \citenamefont
  {Al-Remawi},\ and\ \citenamefont {Woerdman}}]{vanExter1998}%
  \BibitemOpen
  \bibfield  {author} {\bibinfo {author} {\bibfnamefont {M.~P.}\ \bibnamefont
  {{van Exter}}}, \bibinfo {author} {\bibfnamefont {A.}~\bibnamefont
  {Al-Remawi}},\ and\ \bibinfo {author} {\bibfnamefont {J.~P.}\ \bibnamefont
  {Woerdman}},\ }\bibfield  {title} {\bibinfo {title} {Polarization
  fluctuations demonstrate nonlinear anisotropy of a vertical-cavity
  semiconductor laser},\ }\href {https://doi.org/10.1103/PhysRevLett.80.4875}
  {\bibfield  {journal} {\bibinfo  {journal} {Phys. Rev. Lett.}\ }\textbf
  {\bibinfo {volume} {80}},\ \bibinfo {pages} {4875} (\bibinfo {year}
  {1998}{\natexlab{a}})}\BibitemShut {NoStop}%
\bibitem [{\citenamefont {{van Exter}}\ \emph
  {et~al.}(1998{\natexlab{b}})\citenamefont {{van Exter}}, \citenamefont
  {Willemsen},\ and\ \citenamefont {Woerdman}}]{vanExter1998b}%
  \BibitemOpen
  \bibfield  {author} {\bibinfo {author} {\bibfnamefont {M.~P.}\ \bibnamefont
  {{van Exter}}}, \bibinfo {author} {\bibfnamefont {M.~B.}\ \bibnamefont
  {Willemsen}},\ and\ \bibinfo {author} {\bibfnamefont {J.~P.}\ \bibnamefont
  {Woerdman}},\ }\bibfield  {title} {\bibinfo {title} {Polarization
  fluctuations in vertical-cavity semiconductor lasers},\ }\href
  {https://doi.org/10.1103/PhysRevA.58.4191} {\bibfield  {journal} {\bibinfo
  {journal} {Phys. Rev. A}\ }\textbf {\bibinfo {volume} {58}},\ \bibinfo
  {pages} {4191} (\bibinfo {year} {1998}{\natexlab{b}})}\BibitemShut {NoStop}%
\bibitem [{\citenamefont {Sondermann}\ \emph {et~al.}(2004)\citenamefont
  {Sondermann}, \citenamefont {Ackemann}, \citenamefont {Balle}, \citenamefont
  {Mulet},\ and\ \citenamefont {Panajotov}}]{Sondermann2004}%
  \BibitemOpen
  \bibfield  {author} {\bibinfo {author} {\bibfnamefont {M.}~\bibnamefont
  {Sondermann}}, \bibinfo {author} {\bibfnamefont {T.}~\bibnamefont
  {Ackemann}}, \bibinfo {author} {\bibfnamefont {S.}~\bibnamefont {Balle}},
  \bibinfo {author} {\bibfnamefont {J.}~\bibnamefont {Mulet}},\ and\ \bibinfo
  {author} {\bibfnamefont {K.}~\bibnamefont {Panajotov}},\ }\bibfield  {title}
  {\bibinfo {title} {Experimental and theoretical investigations on
  elliptically polarized dynamical transition states in the polarization
  switching of vertical-cavity surface-emitting lasers},\ }\href
  {https://doi.org/https://doi.org/10.1016/j.optcom.2004.02.073} {\bibfield
  {journal} {\bibinfo  {journal} {Opt. Comm.}\ }\textbf {\bibinfo {volume}
  {235}},\ \bibinfo {pages} {421} (\bibinfo {year} {2004})}\BibitemShut
  {NoStop}%
\bibitem [{\citenamefont {Virte}\ \emph
  {et~al.}(2013{\natexlab{b}})\citenamefont {Virte}, \citenamefont
  {Panajotov},\ and\ \citenamefont {Sciamanna}}]{Virte2013a}%
  \BibitemOpen
  \bibfield  {author} {\bibinfo {author} {\bibfnamefont {M.}~\bibnamefont
  {Virte}}, \bibinfo {author} {\bibfnamefont {K.}~\bibnamefont {Panajotov}},\
  and\ \bibinfo {author} {\bibfnamefont {M.}~\bibnamefont {Sciamanna}},\
  }\bibfield  {title} {\bibinfo {title} {Bifurcation to nonlinear polarization
  dynamics and chaos in vertical-cavity surface-emitting lasers},\ }\href
  {https://doi.org/10.1103/PhysRevA.87.013834} {\bibfield  {journal} {\bibinfo
  {journal} {Phys. Rev. A}\ }\textbf {\bibinfo {volume} {87}},\ \bibinfo
  {pages} {013834} (\bibinfo {year} {2013}{\natexlab{b}})}\BibitemShut
  {NoStop}%
\bibitem [{\citenamefont {{San Miguel}}\ \emph {et~al.}(1995)\citenamefont
  {{San Miguel}}, \citenamefont {Feng},\ and\ \citenamefont
  {Moloney}}]{SanMiguel1995}%
  \BibitemOpen
  \bibfield  {author} {\bibinfo {author} {\bibfnamefont {M.}~\bibnamefont {{San
  Miguel}}}, \bibinfo {author} {\bibfnamefont {Q.}~\bibnamefont {Feng}},\ and\
  \bibinfo {author} {\bibfnamefont {J.~V.}\ \bibnamefont {Moloney}},\
  }\bibfield  {title} {\bibinfo {title} {Light-polarization dynamics in
  surface-emitting semiconductor lasers},\ }\href
  {https://doi.org/10.1103/PhysRevA.52.1728} {\bibfield  {journal} {\bibinfo
  {journal} {Phys. Rev. A}\ }\textbf {\bibinfo {volume} {52}},\ \bibinfo
  {pages} {1728} (\bibinfo {year} {1995})}\BibitemShut {NoStop}%
\bibitem [{\citenamefont {Buccafusca}\ \emph {et~al.}(1995)\citenamefont
  {Buccafusca}, \citenamefont {Chilla}, \citenamefont {Rocca}, \citenamefont
  {Wilmsen}, \citenamefont {Feld},\ and\ \citenamefont
  {Leibenguth}}]{Buccafusca1995}%
  \BibitemOpen
  \bibfield  {author} {\bibinfo {author} {\bibfnamefont {O.}~\bibnamefont
  {Buccafusca}}, \bibinfo {author} {\bibfnamefont {J.~L.~A.}\ \bibnamefont
  {Chilla}}, \bibinfo {author} {\bibfnamefont {J.~J.}\ \bibnamefont {Rocca}},
  \bibinfo {author} {\bibfnamefont {C.}~\bibnamefont {Wilmsen}}, \bibinfo
  {author} {\bibfnamefont {S.}~\bibnamefont {Feld}},\ and\ \bibinfo {author}
  {\bibfnamefont {R.}~\bibnamefont {Leibenguth}},\ }\bibfield  {title}
  {\bibinfo {title} {Ultrahigh frequency oscillations and multimode dynamics in
  vertical cavity surface emitting lasers},\ }\href
  {https://doi.org/10.1063/1.114661} {\bibfield  {journal} {\bibinfo  {journal}
  {Appl. Phys. Lett.}\ }\textbf {\bibinfo {volume} {67}},\ \bibinfo {pages}
  {185} (\bibinfo {year} {1995})}\BibitemShut {NoStop}%
\bibitem [{\citenamefont {Buccafusca}\ \emph {et~al.}(1996)\citenamefont
  {Buccafusca}, \citenamefont {Chilla}, \citenamefont {Rocca}, \citenamefont
  {Feld}, \citenamefont {Wilmsen}, \citenamefont {Morozov},\ and\ \citenamefont
  {Leibenguth}}]{Buccafusca1996}%
  \BibitemOpen
  \bibfield  {author} {\bibinfo {author} {\bibfnamefont {O.}~\bibnamefont
  {Buccafusca}}, \bibinfo {author} {\bibfnamefont {J.~L.~A.}\ \bibnamefont
  {Chilla}}, \bibinfo {author} {\bibfnamefont {J.~J.}\ \bibnamefont {Rocca}},
  \bibinfo {author} {\bibfnamefont {S.}~\bibnamefont {Feld}}, \bibinfo {author}
  {\bibfnamefont {C.}~\bibnamefont {Wilmsen}}, \bibinfo {author} {\bibfnamefont
  {V.}~\bibnamefont {Morozov}},\ and\ \bibinfo {author} {\bibfnamefont
  {R.}~\bibnamefont {Leibenguth}},\ }\bibfield  {title} {\bibinfo {title}
  {Transverse mode dynamics in vertical cavity surface emitting lasers excited
  by fast electrical pulses},\ }\href {https://doi.org/10.1063/1.116477}
  {\bibfield  {journal} {\bibinfo  {journal} {Appl. Phys. Lett.}\ }\textbf
  {\bibinfo {volume} {68}},\ \bibinfo {pages} {590} (\bibinfo {year}
  {1996})}\BibitemShut {NoStop}%
\bibitem [{\citenamefont {Buccafusca}\ \emph {et~al.}(1999)\citenamefont
  {Buccafusca}, \citenamefont {Chilla}, \citenamefont {Rocca}, \citenamefont
  {Brusenbach},\ and\ \citenamefont {Martin-Regalado}}]{Buccafusca1999}%
  \BibitemOpen
  \bibfield  {author} {\bibinfo {author} {\bibfnamefont {O.}~\bibnamefont
  {Buccafusca}}, \bibinfo {author} {\bibfnamefont {J.}~\bibnamefont {Chilla}},
  \bibinfo {author} {\bibfnamefont {J.}~\bibnamefont {Rocca}}, \bibinfo
  {author} {\bibfnamefont {P.}~\bibnamefont {Brusenbach}},\ and\ \bibinfo
  {author} {\bibfnamefont {J.}~\bibnamefont {Martin-Regalado}},\ }\bibfield
  {title} {\bibinfo {title} {Transient response of vertical-cavity
  surface-emitting lasers of different active-region diameters},\ }\href
  {https://doi.org/10.1109/3.753666} {\bibfield  {journal} {\bibinfo  {journal}
  {IEEE J. Quant. Electron.}\ }\textbf {\bibinfo {volume} {35}},\ \bibinfo
  {pages} {608} (\bibinfo {year} {1999})}\BibitemShut {NoStop}%
\bibitem [{\citenamefont {Giudici}\ \emph {et~al.}(1998)\citenamefont
  {Giudici}, \citenamefont {Tredicce}, \citenamefont {Vaschenko}, \citenamefont
  {Rocca},\ and\ \citenamefont {Menoni}}]{Giudici1998}%
  \BibitemOpen
  \bibfield  {author} {\bibinfo {author} {\bibfnamefont {M.}~\bibnamefont
  {Giudici}}, \bibinfo {author} {\bibfnamefont {J.}~\bibnamefont {Tredicce}},
  \bibinfo {author} {\bibfnamefont {G.}~\bibnamefont {Vaschenko}}, \bibinfo
  {author} {\bibfnamefont {J.}~\bibnamefont {Rocca}},\ and\ \bibinfo {author}
  {\bibfnamefont {C.}~\bibnamefont {Menoni}},\ }\bibfield  {title} {\bibinfo
  {title} {Spatio-temporal dynamics in vertical cavity surface emitting lasers
  excited by fast electrical pulses},\ }\href
  {https://doi.org/10.1016/s0030-4018(98)00326-5} {\bibfield  {journal}
  {\bibinfo  {journal} {Opt. Comm.}\ }\textbf {\bibinfo {volume} {158}},\
  \bibinfo {pages} {313} (\bibinfo {year} {1998})}\BibitemShut {NoStop}%
\bibitem [{\citenamefont {{Barchanski}}\ \emph {et~al.}(2003)\citenamefont
  {{Barchanski}}, \citenamefont {{Gensty}}, \citenamefont {{Degen}},
  \citenamefont {{Fischer}},\ and\ \citenamefont
  {{Els\"a{\ss}er}}}]{Barchanski2003}%
  \BibitemOpen
  \bibfield  {author} {\bibinfo {author} {\bibfnamefont {A.}~\bibnamefont
  {{Barchanski}}}, \bibinfo {author} {\bibfnamefont {T.}~\bibnamefont
  {{Gensty}}}, \bibinfo {author} {\bibfnamefont {C.}~\bibnamefont {{Degen}}},
  \bibinfo {author} {\bibfnamefont {I.}~\bibnamefont {{Fischer}}},\ and\
  \bibinfo {author} {\bibfnamefont {W.}~\bibnamefont {{Els\"a{\ss}er}}},\
  }\bibfield  {title} {\bibinfo {title} {Picosecond emission dynamics of
  vertical-cavity surface-emitting lasers: spatial, spectral, and
  polarization-resolved characterization},\ }\href
  {https://doi.org/10.1109/JQE.2003.813189} {\bibfield  {journal} {\bibinfo
  {journal} {IEEE J. Quant. Electron.}\ }\textbf {\bibinfo {volume} {39}},\
  \bibinfo {pages} {850} (\bibinfo {year} {2003})}\BibitemShut {NoStop}%
\bibitem [{\citenamefont {Becker}\ \emph {et~al.}(2004)\citenamefont {Becker},
  \citenamefont {Fischer},\ and\ \citenamefont {Els\"a{\ss}er}}]{Becker2004}%
  \BibitemOpen
  \bibfield  {author} {\bibinfo {author} {\bibfnamefont {K.}~\bibnamefont
  {Becker}}, \bibinfo {author} {\bibfnamefont {I.}~\bibnamefont {Fischer}},\
  and\ \bibinfo {author} {\bibfnamefont {W.}~\bibnamefont {Els\"a{\ss}er}},\
  }\bibfield  {title} {\bibinfo {title} {Spatio-temporal emission dynamics of
  vcsels: modal competition in the turn-on behavior},\ }\href
  {https://doi.org/10.1117/12.545883} {\bibfield  {journal} {\bibinfo
  {journal} {Proc. SPIE}\ }\textbf {\bibinfo {volume} {5452}},\ \bibinfo
  {pages} {452} (\bibinfo {year} {2004})}\BibitemShut {NoStop}%
\bibitem [{\citenamefont {Fuchs}\ \emph {et~al.}(2007)\citenamefont {Fuchs},
  \citenamefont {Gensty}, \citenamefont {Debernardi}, \citenamefont {Bava},
  \citenamefont {Ostermann}, \citenamefont {Michalzik}, \citenamefont
  {Haglund}, \citenamefont {Larsson},\ and\ \citenamefont
  {Els\"a{\ss}er}}]{Fuchs2007}%
  \BibitemOpen
  \bibfield  {author} {\bibinfo {author} {\bibfnamefont {C.}~\bibnamefont
  {Fuchs}}, \bibinfo {author} {\bibfnamefont {T.}~\bibnamefont {Gensty}},
  \bibinfo {author} {\bibfnamefont {P.}~\bibnamefont {Debernardi}}, \bibinfo
  {author} {\bibfnamefont {G.~P.}\ \bibnamefont {Bava}}, \bibinfo {author}
  {\bibfnamefont {J.~M.}\ \bibnamefont {Ostermann}}, \bibinfo {author}
  {\bibfnamefont {R.}~\bibnamefont {Michalzik}}, \bibinfo {author}
  {\bibfnamefont {A.}~\bibnamefont {Haglund}}, \bibinfo {author} {\bibfnamefont
  {A.}~\bibnamefont {Larsson}},\ and\ \bibinfo {author} {\bibfnamefont
  {W.}~\bibnamefont {Els\"a{\ss}er}},\ }\bibfield  {title} {\bibinfo {title}
  {Spatiotemporal turn-on dynamics of grating relief vcsels},\ }\href
  {https://doi.org/10.1109/JQE.2007.908118} {\bibfield  {journal} {\bibinfo
  {journal} {IEEE J. Quant. Electron.}\ }\textbf {\bibinfo {volume} {43}},\
  \bibinfo {pages} {1227} (\bibinfo {year} {2007})}\BibitemShut {NoStop}%
\bibitem [{\citenamefont {Richie}\ \emph {et~al.}(1994)\citenamefont {Richie},
  \citenamefont {Zhang}, \citenamefont {Choquette}, \citenamefont {Leibenguth},
  \citenamefont {Zachman},\ and\ \citenamefont {Tabatabaie}}]{Richie1994}%
  \BibitemOpen
  \bibfield  {author} {\bibinfo {author} {\bibfnamefont {D.~A.}\ \bibnamefont
  {Richie}}, \bibinfo {author} {\bibfnamefont {T.}~\bibnamefont {Zhang}},
  \bibinfo {author} {\bibfnamefont {K.~D.}\ \bibnamefont {Choquette}}, \bibinfo
  {author} {\bibfnamefont {R.~E.}\ \bibnamefont {Leibenguth}}, \bibinfo
  {author} {\bibfnamefont {J.~C.}\ \bibnamefont {Zachman}},\ and\ \bibinfo
  {author} {\bibfnamefont {N.}~\bibnamefont {Tabatabaie}},\ }\bibfield  {title}
  {\bibinfo {title} {Chaotic dynamics of mode competition in a vertical-cavity
  surface emitting laser diode under dc excitation},\ }\href
  {https://doi.org/10.1109/3.333701} {\bibfield  {journal} {\bibinfo  {journal}
  {IEEE J. Quant. Electron.}\ }\textbf {\bibinfo {volume} {30}},\ \bibinfo
  {pages} {2500} (\bibinfo {year} {1994})}\BibitemShut {NoStop}%
\bibitem [{\citenamefont {Valle}\ \emph {et~al.}(1995)\citenamefont {Valle},
  \citenamefont {Sarma},\ and\ \citenamefont {Shore}}]{Valle1995}%
  \BibitemOpen
  \bibfield  {author} {\bibinfo {author} {\bibfnamefont {A.}~\bibnamefont
  {Valle}}, \bibinfo {author} {\bibfnamefont {J.}~\bibnamefont {Sarma}},\ and\
  \bibinfo {author} {\bibfnamefont {K.}~\bibnamefont {Shore}},\ }\bibfield
  {title} {\bibinfo {title} {Dynamics of transverse mode competition in
  vertical cavity surface emitting laser diodes},\ }\href
  {https://doi.org/https://doi.org/10.1016/0030-4018(94)00707-2} {\bibfield
  {journal} {\bibinfo  {journal} {Opt. Comm.}\ }\textbf {\bibinfo {volume}
  {115}},\ \bibinfo {pages} {297} (\bibinfo {year} {1995})}\BibitemShut
  {NoStop}%
\bibitem [{\citenamefont {Mart{\'{\i}}n-Regalado}\ \emph
  {et~al.}(1997{\natexlab{a}})\citenamefont {Mart{\'{\i}}n-Regalado},
  \citenamefont {Balle}, \citenamefont {Miguel}, \citenamefont {Valle},\ and\
  \citenamefont {Pesquera}}]{MartinRegalado1997}%
  \BibitemOpen
  \bibfield  {author} {\bibinfo {author} {\bibfnamefont {J.}~\bibnamefont
  {Mart{\'{\i}}n-Regalado}}, \bibinfo {author} {\bibfnamefont {S.}~\bibnamefont
  {Balle}}, \bibinfo {author} {\bibfnamefont {M.~S.}\ \bibnamefont {Miguel}},
  \bibinfo {author} {\bibfnamefont {A.}~\bibnamefont {Valle}},\ and\ \bibinfo
  {author} {\bibfnamefont {L.}~\bibnamefont {Pesquera}},\ }\bibfield  {title}
  {\bibinfo {title} {Polarization and transverse-mode selection in quantum-well
  vertical-cavity surface-emitting lasers: index- and gain-guided devices},\
  }\href {https://doi.org/10.1088/1355-5111/9/5/006} {\bibfield  {journal}
  {\bibinfo  {journal} {Quant. Scl. Opt.}\ }\textbf {\bibinfo {volume} {9}},\
  \bibinfo {pages} {713} (\bibinfo {year} {1997}{\natexlab{a}})}\BibitemShut
  {NoStop}%
\bibitem [{\citenamefont {Mulet}\ and\ \citenamefont
  {Balle}(2002)}]{Mulet2002}%
  \BibitemOpen
  \bibfield  {author} {\bibinfo {author} {\bibfnamefont {J.}~\bibnamefont
  {Mulet}}\ and\ \bibinfo {author} {\bibfnamefont {S.}~\bibnamefont {Balle}},\
  }\bibfield  {title} {\bibinfo {title} {Transverse mode dynamics in
  vertical-cavity surface-emitting lasers: Spatiotemporal versus modal
  expansion descriptions},\ }\href {https://doi.org/10.1103/PhysRevA.66.053802}
  {\bibfield  {journal} {\bibinfo  {journal} {Phys. Rev. A}\ }\textbf {\bibinfo
  {volume} {66}},\ \bibinfo {pages} {053802} (\bibinfo {year}
  {2002})}\BibitemShut {NoStop}%
\bibitem [{\citenamefont {{Mulet}}\ and\ \citenamefont
  {{Balle}}(2002)}]{Mulet2002a}%
  \BibitemOpen
  \bibfield  {author} {\bibinfo {author} {\bibfnamefont {J.}~\bibnamefont
  {{Mulet}}}\ and\ \bibinfo {author} {\bibfnamefont {S.}~\bibnamefont
  {{Balle}}},\ }\bibfield  {title} {\bibinfo {title} {Spatio-temporal modeling
  of the optical properties of vcsels in the presence of polarization
  effects},\ }\href {https://doi.org/10.1109/3.985571} {\bibfield  {journal}
  {\bibinfo  {journal} {IEEE J. Quant. Electron.}\ }\textbf {\bibinfo {volume}
  {38}},\ \bibinfo {pages} {291} (\bibinfo {year} {2002})}\BibitemShut
  {NoStop}%
\bibitem [{\citenamefont {Hess}(1998)}]{Hess1998}%
  \BibitemOpen
  \bibfield  {author} {\bibinfo {author} {\bibfnamefont {O.}~\bibnamefont
  {Hess}},\ }\bibfield  {title} {\bibinfo {title} {Spatio-spectral dynamics and
  spontaneous ultrafast optical switching in vcsel arrays},\ }\href
  {https://doi.org/10.1364/OE.2.000424} {\bibfield  {journal} {\bibinfo
  {journal} {Opt. Express}\ }\textbf {\bibinfo {volume} {2}},\ \bibinfo {pages}
  {424} (\bibinfo {year} {1998})}\BibitemShut {NoStop}%
\bibitem [{\citenamefont {Hess}(2000)}]{Hess2000}%
  \BibitemOpen
  \bibfield  {author} {\bibinfo {author} {\bibfnamefont {O.}~\bibnamefont
  {Hess}},\ }\bibfield  {title} {\bibinfo {title} {Microscopic modelling of
  ultrafast optical switching in coupled arrays of vertical cavity surface
  emitting lasers for optical interconnects},\ }\href
  {https://doi.org/https://doi.org/10.1016/S0030-3992(00)00114-6} {\bibfield
  {journal} {\bibinfo  {journal} {Optics \& Laser Technology}\ }\textbf
  {\bibinfo {volume} {32}},\ \bibinfo {pages} {467} (\bibinfo {year}
  {2000})}\BibitemShut {NoStop}%
\bibitem [{\citenamefont {Babushkin}\ \emph {et~al.}(2011)\citenamefont
  {Babushkin}, \citenamefont {Bandelow},\ and\ \citenamefont
  {Vladimirov}}]{Babushkin2011}%
  \BibitemOpen
  \bibfield  {author} {\bibinfo {author} {\bibfnamefont {I.}~\bibnamefont
  {Babushkin}}, \bibinfo {author} {\bibfnamefont {U.}~\bibnamefont
  {Bandelow}},\ and\ \bibinfo {author} {\bibfnamefont {A.}~\bibnamefont
  {Vladimirov}},\ }\bibfield  {title} {\bibinfo {title} {Rotational symmetry
  breaking in small-area circular vertical cavity surface emitting lasers},\
  }\href {https://doi.org/https://doi.org/10.1016/j.optcom.2010.10.085}
  {\bibfield  {journal} {\bibinfo  {journal} {Opt. Comm.}\ }\textbf {\bibinfo
  {volume} {284}},\ \bibinfo {pages} {1299} (\bibinfo {year}
  {2011})}\BibitemShut {NoStop}%
\bibitem [{\citenamefont {Molitor}\ \emph {et~al.}(2015)\citenamefont
  {Molitor}, \citenamefont {Hartmann},\ and\ \citenamefont
  {Els\"{a}{\ss}er}}]{Molitor2015}%
  \BibitemOpen
  \bibfield  {author} {\bibinfo {author} {\bibfnamefont {A.}~\bibnamefont
  {Molitor}}, \bibinfo {author} {\bibfnamefont {S.}~\bibnamefont {Hartmann}},\
  and\ \bibinfo {author} {\bibfnamefont {W.}~\bibnamefont {Els\"{a}{\ss}er}},\
  }\bibfield  {title} {\bibinfo {title} {Investigations on spatio-spectrally
  resolved stokes polarization parameters of oxide-confined vertical-cavity
  surface-emitting lasers},\ }\href {https://doi.org/10.1364/OL.40.003121}
  {\bibfield  {journal} {\bibinfo  {journal} {Opt. Lett.}\ }\textbf {\bibinfo
  {volume} {40}},\ \bibinfo {pages} {3121} (\bibinfo {year}
  {2015})}\BibitemShut {NoStop}%
\bibitem [{\citenamefont {Mart{\'{\i}}n-Regalado}\ \emph
  {et~al.}(1997{\natexlab{b}})\citenamefont {Mart{\'{\i}}n-Regalado},
  \citenamefont {Chilla}, \citenamefont {Rocca},\ and\ \citenamefont
  {Brusenbach}}]{MartinRegalado1997a}%
  \BibitemOpen
  \bibfield  {author} {\bibinfo {author} {\bibfnamefont {J.}~\bibnamefont
  {Mart{\'{\i}}n-Regalado}}, \bibinfo {author} {\bibfnamefont {J.~L.~A.}\
  \bibnamefont {Chilla}}, \bibinfo {author} {\bibfnamefont {J.~J.}\
  \bibnamefont {Rocca}},\ and\ \bibinfo {author} {\bibfnamefont
  {P.}~\bibnamefont {Brusenbach}},\ }\bibfield  {title} {\bibinfo {title}
  {Polarization switching in vertical-cavity surface emitting lasers observed
  at constant active region temperature},\ }\href
  {https://doi.org/10.1063/1.119167} {\bibfield  {journal} {\bibinfo  {journal}
  {Appl. Phys. Lett.}\ }\textbf {\bibinfo {volume} {70}},\ \bibinfo {pages}
  {3350} (\bibinfo {year} {1997}{\natexlab{b}})}\BibitemShut {NoStop}%
\bibitem [{\citenamefont {Balle}\ \emph {et~al.}(1999)\citenamefont {Balle},
  \citenamefont {Tolkachova}, \citenamefont {Miguel}, \citenamefont {Tredicce},
  \citenamefont {Mart\'{i}n-Regalado},\ and\ \citenamefont {Gahl}}]{Balle1999}%
  \BibitemOpen
  \bibfield  {author} {\bibinfo {author} {\bibfnamefont {S.}~\bibnamefont
  {Balle}}, \bibinfo {author} {\bibfnamefont {E.}~\bibnamefont {Tolkachova}},
  \bibinfo {author} {\bibfnamefont {M.~S.}\ \bibnamefont {Miguel}}, \bibinfo
  {author} {\bibfnamefont {J.~R.}\ \bibnamefont {Tredicce}}, \bibinfo {author}
  {\bibfnamefont {J.}~\bibnamefont {Mart\'{i}n-Regalado}},\ and\ \bibinfo
  {author} {\bibfnamefont {A.}~\bibnamefont {Gahl}},\ }\bibfield  {title}
  {\bibinfo {title} {Mechanisms of polarization switching in
  single-transverse-mode vertical-cavity surface-emitting lasers: thermal shift
  and nonlinear semiconductor dynamics},\ }\href
  {https://doi.org/10.1364/OL.24.001121} {\bibfield  {journal} {\bibinfo
  {journal} {Opt. Lett.}\ }\textbf {\bibinfo {volume} {24}},\ \bibinfo {pages}
  {1121} (\bibinfo {year} {1999})}\BibitemShut {NoStop}%
\bibitem [{\citenamefont {{Olejniczak}}\ \emph {et~al.}(2009)\citenamefont
  {{Olejniczak}}, \citenamefont {{Sciamanna}}, \citenamefont {{Thienpont}},
  \citenamefont {{Panajotov}}, \citenamefont {{Mutig}}, \citenamefont
  {{Hopfer}},\ and\ \citenamefont {{Bimberg}}}]{Olejniczak2009}%
  \BibitemOpen
  \bibfield  {author} {\bibinfo {author} {\bibfnamefont {L.}~\bibnamefont
  {{Olejniczak}}}, \bibinfo {author} {\bibfnamefont {M.}~\bibnamefont
  {{Sciamanna}}}, \bibinfo {author} {\bibfnamefont {H.}~\bibnamefont
  {{Thienpont}}}, \bibinfo {author} {\bibfnamefont {K.}~\bibnamefont
  {{Panajotov}}}, \bibinfo {author} {\bibfnamefont {A.}~\bibnamefont
  {{Mutig}}}, \bibinfo {author} {\bibfnamefont {F.}~\bibnamefont {{Hopfer}}},\
  and\ \bibinfo {author} {\bibfnamefont {D.}~\bibnamefont {{Bimberg}}},\
  }\bibfield  {title} {\bibinfo {title} {Polarization switching in quantum-dot
  vertical-cavity surface-emitting lasers},\ }\href
  {https://doi.org/10.1109/LPT.2009.2021954} {\bibfield  {journal} {\bibinfo
  {journal} {IEEE Phot. Tech. Lett.}\ }\textbf {\bibinfo {volume} {21}},\
  \bibinfo {pages} {1008} (\bibinfo {year} {2009})}\BibitemShut {NoStop}%
\bibitem [{\citenamefont {Degen}\ \emph {et~al.}(1999)\citenamefont {Degen},
  \citenamefont {Fischer},\ and\ \citenamefont {Els\"{a}{\ss}er}}]{Degen1999}%
  \BibitemOpen
  \bibfield  {author} {\bibinfo {author} {\bibfnamefont {C.}~\bibnamefont
  {Degen}}, \bibinfo {author} {\bibfnamefont {I.}~\bibnamefont {Fischer}},\
  and\ \bibinfo {author} {\bibfnamefont {W.}~\bibnamefont {Els\"{a}{\ss}er}},\
  }\bibfield  {title} {\bibinfo {title} {Transverse modes in oxide confined
  vcsels: Influence of pump profile, spatial hole burning, and thermal
  effects},\ }\href {https://doi.org/10.1364/OE.5.000038} {\bibfield  {journal}
  {\bibinfo  {journal} {Opt. Express}\ }\textbf {\bibinfo {volume} {5}},\
  \bibinfo {pages} {38} (\bibinfo {year} {1999})}\BibitemShut {NoStop}%
\bibitem [{\citenamefont {Pereira}\ \emph {et~al.}(1998)\citenamefont
  {Pereira}, \citenamefont {Willemsen}, \citenamefont {van Exter},\ and\
  \citenamefont {Woerdman}}]{Pereira1998}%
  \BibitemOpen
  \bibfield  {author} {\bibinfo {author} {\bibfnamefont {S.~F.}\ \bibnamefont
  {Pereira}}, \bibinfo {author} {\bibfnamefont {M.~B.}\ \bibnamefont
  {Willemsen}}, \bibinfo {author} {\bibfnamefont {M.~P.}\ \bibnamefont {van
  Exter}},\ and\ \bibinfo {author} {\bibfnamefont {J.~P.}\ \bibnamefont
  {Woerdman}},\ }\bibfield  {title} {\bibinfo {title} {Pinning of daisy modes
  in optically pumped vertical-cavity surface-emitting lasers},\ }\href
  {https://doi.org/10.1063/1.121688} {\bibfield  {journal} {\bibinfo  {journal}
  {Appl. Phys. Lett.}\ }\textbf {\bibinfo {volume} {73}},\ \bibinfo {pages}
  {2239} (\bibinfo {year} {1998})}\BibitemShut {NoStop}%
\bibitem [{\citenamefont {{Debernardi}}\ \emph {et~al.}(2002)\citenamefont
  {{Debernardi}}, \citenamefont {{Bava}}, \citenamefont {{Degen}},
  \citenamefont {{Fischer}},\ and\ \citenamefont
  {{Els\"a{\ss}er}}}]{Debernardi2002}%
  \BibitemOpen
  \bibfield  {author} {\bibinfo {author} {\bibfnamefont {P.}~\bibnamefont
  {{Debernardi}}}, \bibinfo {author} {\bibfnamefont {G.~P.}\ \bibnamefont
  {{Bava}}}, \bibinfo {author} {\bibfnamefont {C.}~\bibnamefont {{Degen}}},
  \bibinfo {author} {\bibfnamefont {I.}~\bibnamefont {{Fischer}}},\ and\
  \bibinfo {author} {\bibfnamefont {W.}~\bibnamefont {{Els\"a{\ss}er}}},\
  }\bibfield  {title} {\bibinfo {title} {Influence of anisotropies on
  transverse modes in oxide-confined vcsels},\ }\href
  {https://doi.org/10.1109/3.973322} {\bibfield  {journal} {\bibinfo  {journal}
  {IEEE J. Quant. Electron.}\ }\textbf {\bibinfo {volume} {38}},\ \bibinfo
  {pages} {73} (\bibinfo {year} {2002})}\BibitemShut {NoStop}%
\bibitem [{\citenamefont {Mart\'{i}n-Regalado}\ \emph
  {et~al.}(1997)\citenamefont {Mart\'{i}n-Regalado}, \citenamefont {Balle},\
  and\ \citenamefont {Miguel}}]{MartinRegalado1997c}%
  \BibitemOpen
  \bibfield  {author} {\bibinfo {author} {\bibfnamefont {J.}~\bibnamefont
  {Mart\'{i}n-Regalado}}, \bibinfo {author} {\bibfnamefont {S.}~\bibnamefont
  {Balle}},\ and\ \bibinfo {author} {\bibfnamefont {M.~S.}\ \bibnamefont
  {Miguel}},\ }\bibfield  {title} {\bibinfo {title} {Polarization and
  transverse-mode dynamics of gain-guided vertical-cavity surface-emitting
  lasers},\ }\href {https://doi.org/10.1364/OL.22.000460} {\bibfield  {journal}
  {\bibinfo  {journal} {Opt. Lett.}\ }\textbf {\bibinfo {volume} {22}},\
  \bibinfo {pages} {460} (\bibinfo {year} {1997})}\BibitemShut {NoStop}%
\bibitem [{\citenamefont {Hofmann}\ and\ \citenamefont
  {Hess}(1997)}]{Hofmann1997a}%
  \BibitemOpen
  \bibfield  {author} {\bibinfo {author} {\bibfnamefont {H.~F.}\ \bibnamefont
  {Hofmann}}\ and\ \bibinfo {author} {\bibfnamefont {O.}~\bibnamefont {Hess}},\
  }\bibfield  {title} {\bibinfo {title} {The split-density model: a unified
  description of polarization and array dynamics for vertical-cavity
  surface-emitting lasers},\ }\href {https://doi.org/10.1088/1355-5111/9/5/008}
  {\bibfield  {journal} {\bibinfo  {journal} {Quant. Scl. Opt.}\ }\textbf
  {\bibinfo {volume} {9}},\ \bibinfo {pages} {749} (\bibinfo {year}
  {1997})}\BibitemShut {NoStop}%
\bibitem [{\citenamefont {Molitor}\ \emph {et~al.}(2016)\citenamefont
  {Molitor}, \citenamefont {Mohr}, \citenamefont {Hartmann},\ and\
  \citenamefont {Els\"a{\ss}er}}]{Molitor2016}%
  \BibitemOpen
  \bibfield  {author} {\bibinfo {author} {\bibfnamefont {A.}~\bibnamefont
  {Molitor}}, \bibinfo {author} {\bibfnamefont {T.}~\bibnamefont {Mohr}},
  \bibinfo {author} {\bibfnamefont {S.}~\bibnamefont {Hartmann}},\ and\
  \bibinfo {author} {\bibfnamefont {W.}~\bibnamefont {Els\"a{\ss}er}},\
  }\bibfield  {title} {\bibinfo {title} {Investigations on spectro-temporally
  resolved stokes polarization parameters of transverse multi-mode
  vertical-cavity surface-emitting lasers},\ }\href
  {https://doi.org/10.1109/JQE.2015.2505506} {\bibfield  {journal} {\bibinfo
  {journal} {IEEE J. Quant. Electron.}\ }\textbf {\bibinfo {volume} {52}},\
  \bibinfo {pages} {2400109} (\bibinfo {year} {2016})}\BibitemShut {NoStop}%
\bibitem [{\citenamefont {Pl\"oschner}\ \emph {et~al.}(2022)\citenamefont
  {Pl\"oschner}, \citenamefont {Morote}, \citenamefont {Dahl}, \citenamefont
  {Mounaix}, \citenamefont {Light}, \citenamefont {Raki\'c},\ and\
  \citenamefont {Carpenter}}]{Ploeschner2022}%
  \BibitemOpen
  \bibfield  {author} {\bibinfo {author} {\bibfnamefont {M.}~\bibnamefont
  {Pl\"oschner}}, \bibinfo {author} {\bibfnamefont {M.~M.}\ \bibnamefont
  {Morote}}, \bibinfo {author} {\bibfnamefont {D.~S.}\ \bibnamefont {Dahl}},
  \bibinfo {author} {\bibfnamefont {M.}~\bibnamefont {Mounaix}}, \bibinfo
  {author} {\bibfnamefont {G.}~\bibnamefont {Light}}, \bibinfo {author}
  {\bibfnamefont {A.}~\bibnamefont {Raki\'c}},\ and\ \bibinfo {author}
  {\bibfnamefont {J.}~\bibnamefont {Carpenter}},\ }\bibfield  {title} {\bibinfo
  {title} {Spatial tomography of light resolved in time, spectrum, and
  polarisation},\ }\href {https://doi.org/10.1038/s41467-022-31814-2}
  {\bibfield  {journal} {\bibinfo  {journal} {Nature Comm.}\ }\textbf {\bibinfo
  {volume} {13}},\ \bibinfo {pages} {4294} (\bibinfo {year}
  {2022})}\BibitemShut {NoStop}%
\bibitem [{\citenamefont {Bouchez}\ \emph {et~al.}(2019)\citenamefont
  {Bouchez}, \citenamefont {Uy}, \citenamefont {Macias}, \citenamefont
  {Wolfersberger},\ and\ \citenamefont {Sciamanna}}]{Bouchez2019}%
  \BibitemOpen
  \bibfield  {author} {\bibinfo {author} {\bibfnamefont {G.}~\bibnamefont
  {Bouchez}}, \bibinfo {author} {\bibfnamefont {C.-H.}\ \bibnamefont {Uy}},
  \bibinfo {author} {\bibfnamefont {B.}~\bibnamefont {Macias}}, \bibinfo
  {author} {\bibfnamefont {D.}~\bibnamefont {Wolfersberger}},\ and\ \bibinfo
  {author} {\bibfnamefont {M.}~\bibnamefont {Sciamanna}},\ }\bibfield  {title}
  {\bibinfo {title} {Wideband chaos from a laser diode with phase-conjugate
  feedback},\ }\href {https://doi.org/10.1364/OL.44.000975} {\bibfield
  {journal} {\bibinfo  {journal} {Opt. Lett.}\ }\textbf {\bibinfo {volume}
  {44}},\ \bibinfo {pages} {975} (\bibinfo {year} {2019})}\BibitemShut
  {NoStop}%
\bibitem [{\citenamefont {Porte}\ \emph {et~al.}(2021)\citenamefont {Porte},
  \citenamefont {Skalli}, \citenamefont {Haghighi}, \citenamefont
  {Reitzenstein}, \citenamefont {Lott},\ and\ \citenamefont
  {Brunner}}]{Porte2021}%
  \BibitemOpen
  \bibfield  {author} {\bibinfo {author} {\bibfnamefont {X.}~\bibnamefont
  {Porte}}, \bibinfo {author} {\bibfnamefont {A.}~\bibnamefont {Skalli}},
  \bibinfo {author} {\bibfnamefont {N.}~\bibnamefont {Haghighi}}, \bibinfo
  {author} {\bibfnamefont {S.}~\bibnamefont {Reitzenstein}}, \bibinfo {author}
  {\bibfnamefont {J.~A.}\ \bibnamefont {Lott}},\ and\ \bibinfo {author}
  {\bibfnamefont {D.}~\bibnamefont {Brunner}},\ }\bibfield  {title} {\bibinfo
  {title} {A complete, parallel and autonomous photonic neural network in a
  semiconductor multimode laser},\ }\href
  {https://doi.org/10.1088/2515-7647/abf6bd} {\bibfield  {journal} {\bibinfo
  {journal} {J. Phys. Phot.}\ }\textbf {\bibinfo {volume} {3}},\ \bibinfo
  {pages} {024017} (\bibinfo {year} {2021})}\BibitemShut {NoStop}%
\bibitem [{\citenamefont {Vatin}\ \emph {et~al.}(2019)\citenamefont {Vatin},
  \citenamefont {Rontani},\ and\ \citenamefont {Sciamanna}}]{Vatin2019}%
  \BibitemOpen
  \bibfield  {author} {\bibinfo {author} {\bibfnamefont {J.}~\bibnamefont
  {Vatin}}, \bibinfo {author} {\bibfnamefont {D.}~\bibnamefont {Rontani}},\
  and\ \bibinfo {author} {\bibfnamefont {M.}~\bibnamefont {Sciamanna}},\
  }\bibfield  {title} {\bibinfo {title} {Experimental reservoir computing using
  vcsel polarization dynamics},\ }\href {https://doi.org/10.1364/OE.27.018579}
  {\bibfield  {journal} {\bibinfo  {journal} {Opt. Express}\ }\textbf {\bibinfo
  {volume} {27}},\ \bibinfo {pages} {18579} (\bibinfo {year}
  {2019})}\BibitemShut {NoStop}%
\bibitem [{\citenamefont {Chen}\ \emph {et~al.}(2007)\citenamefont {Chen},
  \citenamefont {Su}, \citenamefont {Lu}, \citenamefont {Liu}, \citenamefont
  {Chen},\ and\ \citenamefont {Huang}}]{Chen2007b}%
  \BibitemOpen
  \bibfield  {author} {\bibinfo {author} {\bibfnamefont {C.~C.}\ \bibnamefont
  {Chen}}, \bibinfo {author} {\bibfnamefont {K.~W.}\ \bibnamefont {Su}},
  \bibinfo {author} {\bibfnamefont {T.~H.}\ \bibnamefont {Lu}}, \bibinfo
  {author} {\bibfnamefont {C.~C.}\ \bibnamefont {Liu}}, \bibinfo {author}
  {\bibfnamefont {Y.~F.}\ \bibnamefont {Chen}},\ and\ \bibinfo {author}
  {\bibfnamefont {K.~F.}\ \bibnamefont {Huang}},\ }\bibfield  {title} {\bibinfo
  {title} {Generation of two-dimensional chaotic vector fields from a
  surface-emitting semiconductor laser: Analysis of vector singularities},\
  }\href {https://doi.org/10.1103/PhysRevE.76.026219} {\bibfield  {journal}
  {\bibinfo  {journal} {Phys. Rev. E}\ }\textbf {\bibinfo {volume} {76}},\
  \bibinfo {pages} {026219} (\bibinfo {year} {2007})}\BibitemShut {NoStop}%
\bibitem [{\citenamefont {Babushkin}\ \emph {et~al.}(2008)\citenamefont
  {Babushkin}, \citenamefont {Schulz-Ruhtenberg}, \citenamefont {Loiko},
  \citenamefont {Huang},\ and\ \citenamefont {Ackemann}}]{Babushkin2008}%
  \BibitemOpen
  \bibfield  {author} {\bibinfo {author} {\bibfnamefont {I.~V.}\ \bibnamefont
  {Babushkin}}, \bibinfo {author} {\bibfnamefont {M.}~\bibnamefont
  {Schulz-Ruhtenberg}}, \bibinfo {author} {\bibfnamefont {N.~A.}\ \bibnamefont
  {Loiko}}, \bibinfo {author} {\bibfnamefont {K.~F.}\ \bibnamefont {Huang}},\
  and\ \bibinfo {author} {\bibfnamefont {T.}~\bibnamefont {Ackemann}},\
  }\bibfield  {title} {\bibinfo {title} {Coupling of polarization and spatial
  degrees of freedom of highly divergent emission in broad-area square
  vertical-cavity surface-emitting lasers},\ }\href
  {https://doi.org/10.1103/PhysRevLett.100.213901} {\bibfield  {journal}
  {\bibinfo  {journal} {Phys. Rev. Lett.}\ }\textbf {\bibinfo {volume} {100}},\
  \bibinfo {pages} {213901} (\bibinfo {year} {2008})}\BibitemShut {NoStop}%
\bibitem [{\citenamefont {Huang}\ \emph {et~al.}(2002)\citenamefont {Huang},
  \citenamefont {Chen}, \citenamefont {Lai},\ and\ \citenamefont
  {Lan}}]{Huang2002}%
  \BibitemOpen
  \bibfield  {author} {\bibinfo {author} {\bibfnamefont {K.~F.}\ \bibnamefont
  {Huang}}, \bibinfo {author} {\bibfnamefont {Y.~F.}\ \bibnamefont {Chen}},
  \bibinfo {author} {\bibfnamefont {H.~C.}\ \bibnamefont {Lai}},\ and\ \bibinfo
  {author} {\bibfnamefont {Y.~P.}\ \bibnamefont {Lan}},\ }\bibfield  {title}
  {\bibinfo {title} {Observation of the wave function of a quantum billiard
  from the transverse patterns of vertical cavity surface emitting lasers},\
  }\href {https://doi.org/10.1103/PhysRevLett.89.224102} {\bibfield  {journal}
  {\bibinfo  {journal} {Phys. Rev. Lett.}\ }\textbf {\bibinfo {volume} {89}},\
  \bibinfo {pages} {224102} (\bibinfo {year} {2002})}\BibitemShut {NoStop}%
\bibitem [{\citenamefont {Chen}\ \emph {et~al.}(2003)\citenamefont {Chen},
  \citenamefont {Huang}, \citenamefont {Lai},\ and\ \citenamefont
  {Lan}}]{Chen2003a}%
  \BibitemOpen
  \bibfield  {author} {\bibinfo {author} {\bibfnamefont {Y.~F.}\ \bibnamefont
  {Chen}}, \bibinfo {author} {\bibfnamefont {K.~F.}\ \bibnamefont {Huang}},
  \bibinfo {author} {\bibfnamefont {H.~C.}\ \bibnamefont {Lai}},\ and\ \bibinfo
  {author} {\bibfnamefont {Y.~P.}\ \bibnamefont {Lan}},\ }\bibfield  {title}
  {\bibinfo {title} {Observation of vector vortex lattices in polarization
  states of an isotropic microcavity laser},\ }\href
  {https://doi.org/10.1103/PhysRevLett.90.053904} {\bibfield  {journal}
  {\bibinfo  {journal} {Phys. Rev. Lett.}\ }\textbf {\bibinfo {volume} {90}},\
  \bibinfo {pages} {053904} (\bibinfo {year} {2003})}\BibitemShut {NoStop}%
\bibitem [{\citenamefont {Brejnak}\ \emph {et~al.}(2021)\citenamefont
  {Brejnak}, \citenamefont {Gebski}, \citenamefont {Sok{\'{o}}{\l}},
  \citenamefont {Marciniak}, \citenamefont {Wasiak}, \citenamefont {Muszalski},
  \citenamefont {Lott}, \citenamefont {Fischer},\ and\ \citenamefont
  {Czyszanowski}}]{Brejnak2021}%
  \BibitemOpen
  \bibfield  {author} {\bibinfo {author} {\bibfnamefont {A.}~\bibnamefont
  {Brejnak}}, \bibinfo {author} {\bibfnamefont {M.}~\bibnamefont {Gebski}},
  \bibinfo {author} {\bibfnamefont {A.~K.}\ \bibnamefont {Sok{\'{o}}{\l}}},
  \bibinfo {author} {\bibfnamefont {M.}~\bibnamefont {Marciniak}}, \bibinfo
  {author} {\bibfnamefont {M.}~\bibnamefont {Wasiak}}, \bibinfo {author}
  {\bibfnamefont {J.}~\bibnamefont {Muszalski}}, \bibinfo {author}
  {\bibfnamefont {J.~A.}\ \bibnamefont {Lott}}, \bibinfo {author}
  {\bibfnamefont {I.}~\bibnamefont {Fischer}},\ and\ \bibinfo {author}
  {\bibfnamefont {T.}~\bibnamefont {Czyszanowski}},\ }\bibfield  {title}
  {\bibinfo {title} {Boosting the output power of large-aperture lasers by
  breaking their circular symmetry},\ }\href
  {https://doi.org/10.1364/optica.421753} {\bibfield  {journal} {\bibinfo
  {journal} {Optica}\ }\textbf {\bibinfo {volume} {8}},\ \bibinfo {pages}
  {1167} (\bibinfo {year} {2021})}\BibitemShut {NoStop}%
\bibitem [{\citenamefont {Molitor}\ \emph {et~al.}(2012)\citenamefont
  {Molitor}, \citenamefont {Hartmann},\ and\ \citenamefont
  {Els\"{a}{\ss}er}}]{Molitor2012}%
  \BibitemOpen
  \bibfield  {author} {\bibinfo {author} {\bibfnamefont {A.}~\bibnamefont
  {Molitor}}, \bibinfo {author} {\bibfnamefont {S.}~\bibnamefont {Hartmann}},\
  and\ \bibinfo {author} {\bibfnamefont {W.}~\bibnamefont {Els\"{a}{\ss}er}},\
  }\bibfield  {title} {\bibinfo {title} {Stokes vector characterization of the
  polarization behavior of vertical-cavity surface-emitting lasers},\ }\href
  {https://doi.org/10.1364/OL.37.004799} {\bibfield  {journal} {\bibinfo
  {journal} {Opt. Lett.}\ }\textbf {\bibinfo {volume} {37}},\ \bibinfo {pages}
  {4799} (\bibinfo {year} {2012})}\BibitemShut {NoStop}%
\bibitem [{\citenamefont {Chang-Hasnain}\ \emph {et~al.}(1991)\citenamefont
  {Chang-Hasnain}, \citenamefont {Harbison}, \citenamefont {Hasnain},
  \citenamefont {Von~Lehmen}, \citenamefont {Florez},\ and\ \citenamefont
  {Stoffel}}]{ChangHasnain1991}%
  \BibitemOpen
  \bibfield  {author} {\bibinfo {author} {\bibfnamefont {C.}~\bibnamefont
  {Chang-Hasnain}}, \bibinfo {author} {\bibfnamefont {J.}~\bibnamefont
  {Harbison}}, \bibinfo {author} {\bibfnamefont {G.}~\bibnamefont {Hasnain}},
  \bibinfo {author} {\bibfnamefont {A.}~\bibnamefont {Von~Lehmen}}, \bibinfo
  {author} {\bibfnamefont {L.}~\bibnamefont {Florez}},\ and\ \bibinfo {author}
  {\bibfnamefont {N.}~\bibnamefont {Stoffel}},\ }\bibfield  {title} {\bibinfo
  {title} {Dynamic, polarization, and transverse mode characteristics of
  vertical cavity surface emitting lasers},\ }\href
  {https://doi.org/10.1109/3.89957} {\bibfield  {journal} {\bibinfo  {journal}
  {IEEE J. Quant. Electron.}\ }\textbf {\bibinfo {volume} {27}},\ \bibinfo
  {pages} {1402} (\bibinfo {year} {1991})}\BibitemShut {NoStop}%
\bibitem [{\citenamefont {Degen}\ \emph {et~al.}(2000)\citenamefont {Degen},
  \citenamefont {Krauskopf}, \citenamefont {Jennemann}, \citenamefont
  {Fischer},\ and\ \citenamefont {Els\"a{\ss}er}}]{Degen2000}%
  \BibitemOpen
  \bibfield  {author} {\bibinfo {author} {\bibfnamefont {C.}~\bibnamefont
  {Degen}}, \bibinfo {author} {\bibfnamefont {B.}~\bibnamefont {Krauskopf}},
  \bibinfo {author} {\bibfnamefont {G.}~\bibnamefont {Jennemann}}, \bibinfo
  {author} {\bibfnamefont {I.}~\bibnamefont {Fischer}},\ and\ \bibinfo {author}
  {\bibfnamefont {W.}~\bibnamefont {Els\"a{\ss}er}},\ }\bibfield  {title}
  {\bibinfo {title} {Polarization selective symmetry breaking in the
  near-fields of vertical cavity surface emitting lasers},\ }\href
  {https://doi.org/10.1088/1464-4266/2/4/310} {\bibfield  {journal} {\bibinfo
  {journal} {Quant. Scl. Opt.}\ }\textbf {\bibinfo {volume} {2}},\ \bibinfo
  {pages} {517} (\bibinfo {year} {2000})}\BibitemShut {NoStop}%
\bibitem [{\citenamefont {Misak}\ \emph {et~al.}(2015)\citenamefont {Misak},
  \citenamefont {Dugmore}, \citenamefont {Middleton}, \citenamefont {Hale},
  \citenamefont {Farner}, \citenamefont {Choquette},\ and\ \citenamefont
  {Leisher}}]{Misak2015}%
  \BibitemOpen
  \bibfield  {author} {\bibinfo {author} {\bibfnamefont {S.~M.}\ \bibnamefont
  {Misak}}, \bibinfo {author} {\bibfnamefont {D.~G.}\ \bibnamefont {Dugmore}},
  \bibinfo {author} {\bibfnamefont {K.~A.}\ \bibnamefont {Middleton}}, \bibinfo
  {author} {\bibfnamefont {E.~R.}\ \bibnamefont {Hale}}, \bibinfo {author}
  {\bibfnamefont {K.~R.}\ \bibnamefont {Farner}}, \bibinfo {author}
  {\bibfnamefont {K.~D.}\ \bibnamefont {Choquette}},\ and\ \bibinfo {author}
  {\bibfnamefont {P.~O.}\ \bibnamefont {Leisher}},\ }\bibfield  {title}
  {\bibinfo {title} {Spectrally resolved imaging of the transverse modes in
  multimode vcsels},\ }\href {https://doi.org/10.1117/12.2076629} {\bibfield
  {journal} {\bibinfo  {journal} {Proc. SPIE}\ }\textbf {\bibinfo {volume}
  {9381}},\ \bibinfo {pages} {93810L} (\bibinfo {year} {2015})}\BibitemShut
  {NoStop}%
\end{thebibliography}
\end{document}